\documentclass[%
 reprint,
 superscriptaddress,
 preprintnumbers,
 amsmath,amssymb,
 prb,
floatfix,
]{revtex4-2}

\usepackage[utf8x]{inputenc}
\usepackage[final]{changes}
\usepackage{graphicx}
\usepackage{braket}
\usepackage{verbatim}
\usepackage{hyperref}
\usepackage{subfigure}
\usepackage{longtable}
\usepackage{tabularx}
\usepackage{dcolumn}%
\usepackage{color}
\usepackage[normalem]{ulem}

\newcommand{\e}{\ensuremath{\mathrm{e}}}
\newcommand{\iu}{\ensuremath{\mathrm{i}}}
\newcommand{\dd}{\mathrm{d}}

\begin{document}

\author{Stefano Paolo Villani}
\affiliation{Dipartimento di Fisica, Università di Roma La Sapienza, Rome, Italy}

\author{Marco Campetella}
\affiliation{CNR-SPIN Institute for Superconducting and other Innovative Materials and Devices, Area della Ricerca di Tor Vergata,
Via del Fosso del Cavaliere 100, I-00133 Rome, Italy}

\author{Paolo Barone}
\affiliation{CNR-SPIN Institute for Superconducting and other Innovative Materials and Devices, Area della Ricerca di Tor Vergata,
Via del Fosso del Cavaliere 100, I-00133 Rome, Italy}

\author{Francesco Mauri}
\email[]{francesco.mauri@uniroma1.it}
\affiliation{Dipartimento di Fisica, Università di Roma La Sapienza, Rome, Italy}

\title{Giant piezoelectricity driven by Thouless pump in conjugated polymers}

\begin{abstract}
Piezoelectricity of organic polymers has attracted increasing interest because of several advantages they exhibit over traditional inorganic ceramics. While most organic piezoelectrics rely on the presence of intrinsic local dipoles, a highly nonlocal electronic polarization can be foreseen in conjugated polymers, characterised by delocalized and highly responsive $\pi$-electrons. These 1D systems represent a physical realization of a Thouless pump, a mechanism of adiabatic charge transport of topological nature which results, as shown in this work, in anomalously large dynamical effective charges, inversely proportional to the band gap energy. A structural (ferroelectric) phase transition further contributes to an enhancement of the piezoelectric response reminiscent of that observed in piezoelectric perovskites close to morphotropic phase boundaries. First-principles Density Functional Theory (DFT) calculations performed in two representative conjugated polymers using hybrid functionals, show that state-of-the-art organic piezoelectric are outperformed by piezoelectric conjugated polymers, mostly thanks to strongly anomalous effective charges of carbon, larger than 5$e$ -- ordinary values being of the order of 1$e$ -- and reaching the giant value of 30$e$ for band gaps of the order of 1 eV.
\end{abstract}

\maketitle

\section*{Introduction}
Piezoelectricity is a well known phenomenon characterising those materials which have the property to generate a surface charge, and hence an electric tension, when subject to a stress, or conversely to deform elastically in response to an external electric field. Thanks to the possibility they offer to convert mechanical energy into electrical energy and vice-versa, piezoelectric materials are of great interest in various fields and for many applications, from macro- to microscopic electromechanical devices, to energy harvesting and much more\cite{tadigadapa2009piezoelectric,briscoe2015piezoelectric}. The most widely used piezoelectric materials are inorganic perovskites, such as lead zirconate titanate (PZT), for their high electromechanical response\cite{berlincourt1971piezoelectric,jaffe1958piezoelectric,damjanovic2009review}. However, inorganic materials have very low mechanical flexibility, high fabrication costs and often are toxic because of the lead they contain. These facts motivated an intense research activity aimed at developing and identifying lead-free piezoelectric ceramics\cite{damjanovic2009review,saito2004review}. A promising alternative is to exploit piezoelectric properties of organic materials, a route that has been trodden with some success since early 1990s, mostly focusing on the wide class of organic polymers displaying high flexibility, low fabrication costs and bio-compatibility\cite{setter2006ferroelectric,lovinger1983ferroelectric,ramadan2014review,bowen2022advfuncmat}. In this context, the most studied organic piezoelectric is polyvinylidene fluoride (PVDF)\cite{lovinger1983ferroelectric,PVDF_review2020}, a saturated polymer derived from polyethylene and comprising molecular units with net (electrical) dipole moments,  thus giving rise to ferroelectricity of conformational origin due to the rotation of chains' segments from non-polar to polar isomers.

Despite the intense research efforts and the progress made in the field of organic piezoelectrics, to date inorganic ceramics still display much better piezoelectric performance than organic counterparts. The piezoelectric response in PZT and related inorganic materials is strongly enhanced at morphotropic phase boundaries (MPB), marking a composition-driven structural transition between two competing, nearly energetically degenerate phases with distinct symmetries\cite{2000nature.fu,2006nature.kutnjak,ahart2008origin}. On general grounds, the properties enhancement close to such phase transition may be traced back to the flattening of free-energy surfaces, easing polarization extension and/or rotation, as extensively discussed and observed mostly in perovskite oxides\cite{damjanovic2010morphotropic}. Recently, the concept of MPB has been loosely extended to the family of P(VDF-TrFE) copolymers\cite{2018liu.nature,small2022}. Here, the introduction of different TrFE monomers in the semicrystalline PVDF structure has been proposed to lead to an enhanced conformational competition reminiscent of the structural competition realized at MPB, further suggesting an optimal chemical composition for maximizing the piezoelectric coefficient. Even though at the optimal "morphotropic" composition the piezoelectric coefficient roughly doubles the typical values of PVDF, it is still smaller by one order of magnitude compared to characteristic piezoelectric coefficients of inorganic oxide ceramics such as PZT.

Beside saturated polymers as PVDF and P(VDF-TrFE), whose ferroelectric and piezoelectric properties rely on the ordering of built-in molecular dipoles and as such require appropriate poling treatments, a natural alternative choice is represented by conjugated polymers, such as archetypical polyacetylene (PA). 
Characterised by a delocalised $\pi-$orbital along their backbone, these polymers are widely studied for their peculiar electronic properties\cite{barford2013electronic,bronstein2020role}.  If specific symmetry-lowering effects allowing for piezoelectricity and ferroelectricity are met in a conjugated polymer, one may expect a large polar response of electronic origin due to the redistribution of the responsive $\pi-$electronic density along the polymer's backbone. Because of the delocalized nature of conjugated $\pi-$state (of Bloch-type in an infinite periodic chain), small changes of atomic positions may lead to a global shift of electrons, revealing the strong nonlocal and ultimately topological character of electronic polarization in these quasi-1D systems\cite{onoda2004topological}. Indeed, conjugated polymers have been theoretically put forward as a potential new class of ``electronic ferroelectrics''\cite{KIROVA2016}.

In order to explore the potential piezoelectricity of conjugated polymers, we adopt the simple model originally introduced by Rice and Mele to study soliton excitations of linearly conjugated diatomic polymers\cite{rice1982elementary}, later become a prototypical model for 1D ferroelectricity\cite{onoda2004topological,yamauchi2014electronic} and Thouless adiabatic pumping\cite{vanderbilt1993electric,xiao2010berry}. Building on these well-established results, we study how the interplay between these mechanisms affect the system response to a mechanical strain. 
We find that polar responses, such as Born effective-charge tensors -- also known as atomic polar tensors -- and piezoelectric coefficients, can be strongly enhanced close to the dimerization phase transition characterized by a bond-length alternation (BLA) between neighboring atoms along the chain. This transition can be controlled by changes in the chemical composition of the polymers or by tuning the electron-phonon (e-ph) interaction, that is expected to be sensitive to the dielectric environment of individual chains.  The strength of the piezoelectric effect is found to be determined partly by the MPB-like enhancement -- relying on the second-order nature of the transition --, but mostly by a giant effective charge, ultimately due to the topological nature of Thouless' adiabatic charge transport in this class of materials. We verified the model-based predictions and theoretical picture by performing Density Functional Theory (DFT) calculations using a range-separated hybrid functional on two prototypical conjugated polymers derived from polyacetylene, confirming strong enhancement of effective charges and consequently of piezoelectric response, that is found to be larger than the one observed in the celebrated organic piezoelectric PVDF for a wide range of parameters.

\section*{Results}
\subsection{A model for piezoelectric conjugated polymers}\label{sec:model}
To study the effects of strain on conjugated polymers, we generalise the Rice-Mele model\cite{rice1982elementary} of a 1D linear chain with a unit cell of length $a_0$ containing two atoms. The only possible strains for a 1D infinite periodic chain are contraction/dilatation of the unit cell, defined by $a(\epsilon) = a_0(1+\epsilon)$, where the adimensional parameter $\epsilon$ quantifies the strain along the direction of the chain. In a nearest neighbours tight-binding approximation, the modulation of the hopping energy between atoms at site $i$ and $i+1$ can be modelled at linear order in atoms' displacement and strain as $t_{i,i+1} = t(\epsilon) + (-1)^{i+1} \delta t(\epsilon)$, with
\begin{align}
    \label{eq:def_t^eq(eps)}
    t(\epsilon) &= t_0(1-\beta\epsilon), \\
    \label{eq:def_dt(eps)}
    \delta t(\epsilon) &= -t_0\beta (1+\epsilon) u(\epsilon). 
\end{align}
Equation (\ref{eq:def_t^eq(eps)}) accounts for the effect of strain on the hopping energy $t_0$ between equidistant atoms, while Equation (\ref{eq:def_dt(eps)}) describes the variation with respect to $t(\epsilon)$ caused by atoms' displacement; at the lowest order, we can assume the same adimensional e-ph coupling parameter $\beta>0$ in describing both effects. The term $u(\epsilon)$ is an adimensional fractional coordinate measuring the relative displacement of the two atoms in the unit cell, as defined in the Methods. We indicate with the term $\Delta\geq0$ the on-site energy difference between neighbours. If $\Delta=0$, the atoms in the unit cell are equivalent and we recover the well-known SSH model\cite{su1979solitons}, used to describe polyacetylene. However, in order to have a non-trivial polar response, it is necessary to break atoms' equivalence ($\Delta\neq0$) as, e.g., in substituted polyacetylenes (SPA)\cite{masuda2007substituted}, a class of conjugated polymers formed by inequivalent monomers which can be obtained substituting (one of) the atoms in the C$_2$H$_2$ unit of PA with some element(s) or compound(s). A representation of the model is in the insets of Figure \ref{fig:model_properties}a) and \ref{fig:model_properties}b).
For simplicity, we consider only longitudinal displacements, parallel to the linear-chain direction. As we are interested in the equilibrium structures at $T= 0~K$, we define the optimal displacement $\overline{u}$ as the one which minimises the total energy, given a set of material-dependent parameters and a strain. For more information on the model and the derivation of the above quantities, see the Methods and the Supplementary Information.

In the absence of strain ($\epsilon=0$), the properties of the model are well known. In the SSH model ($\Delta=0$) a finite $\overline{u}\neq0$ is allowed by an infinitely small e-ph interaction, conjuring with a Fermi-surface nesting and Peierls electronic instability to produce a dimerized phase with bond-length alternation and the opening of a gap in the energy spectrum $E_{\mathrm{gap}} = 4|\delta t| = 4 \beta t_0\overline{u}$. Breaking the equivalence between atoms ($\Delta\neq0$) also opens a gap which is expected to counteract the Peierls instability and the related bond dimerization. The lower-symmetry structure can be stabilised only by a finite and strong enough e-ph interaction, resulting in a gap $E_{\mathrm{gap}} = 2\sqrt{\Delta^2 + 4 \delta t^2} = 2\sqrt{\Delta^2 + 4 (\beta t_0\overline{u})^2}$. Indeed, as shown in Figure \ref{fig:model_properties}a), increasing $\Delta$ at fixed e-ph coupling induces a second order structural phase transition at a critical value $\Delta_{\mathrm{c}}$ with order parameter $\overline{u}$ between a distorted ($\overline{u}\neq0$) and a higher-symmetric undimerized ($\overline{u}=0$) phase, both displaying an insulating character with a finite band-gap (see Supplementary Information). Such structural transition from the dimerized to undimerized phase can also be induced by decreasing the e-ph coupling at fixed non-zero onsite energy difference, as discussed in the next section and displayed in Figure \ref{fig:model_properties}b). In a 2D parametric ($\Delta,\delta t$)-space, where $\delta t$ depends -- through optimal $\overline{u}$ -- on both $\Delta$ and $\beta$, the origin of the axes corresponds to a metallic system while the rest of the space to an insulator. Since the milestone paper by Vanderbilt and King-Smith \cite{vanderbilt1993electric}, it is known that if this system undergoes an adiabatic evolution along a loop enclosing the origin of this space, a quantised charge $|e|$ is pumped out. This is a prototypical example of the adiabatic charge transport known as Thouless' pump\cite{thouless1983quantization,xiao2010berry} realised e.g. in ultracold fermions\cite{nakajima2016topological}. The crucial ingredient is the presence of the metallic point in the domain enclosed by the loop and, in this sense, it is an example of topological phenomenon (see Supplementary Information). Previous studies on low-dimensional systems showed how topology may reflect onto polar responses, resulting, e.g., in electronic polarization in quasi-1D systems\cite{onoda2004topological} or in large effective charges and piezoelectric coefficients directly related to valley Chern numbers in 2D systems\cite{rostami2018npj,bistoni2019giant}. In the next two Sections, we use the generalized Rice-Mele model to show how the Thouless-pump topological mechanism conjures with the bond dimerization to give raise to an enhanced piezoelectric effect, and we discuss how the latter can be tuned by acting on the two independent parameters $\Delta$ and $\beta$.

\subsection{Morphotropic-like enhancement of the piezoelectric response}\label{sec:MPB_enhancement}
\begin{figure*}[!htb]
    \begin{minipage}[c]{1.0\linewidth}
    \centering
    \includegraphics[width=0.475\textwidth]{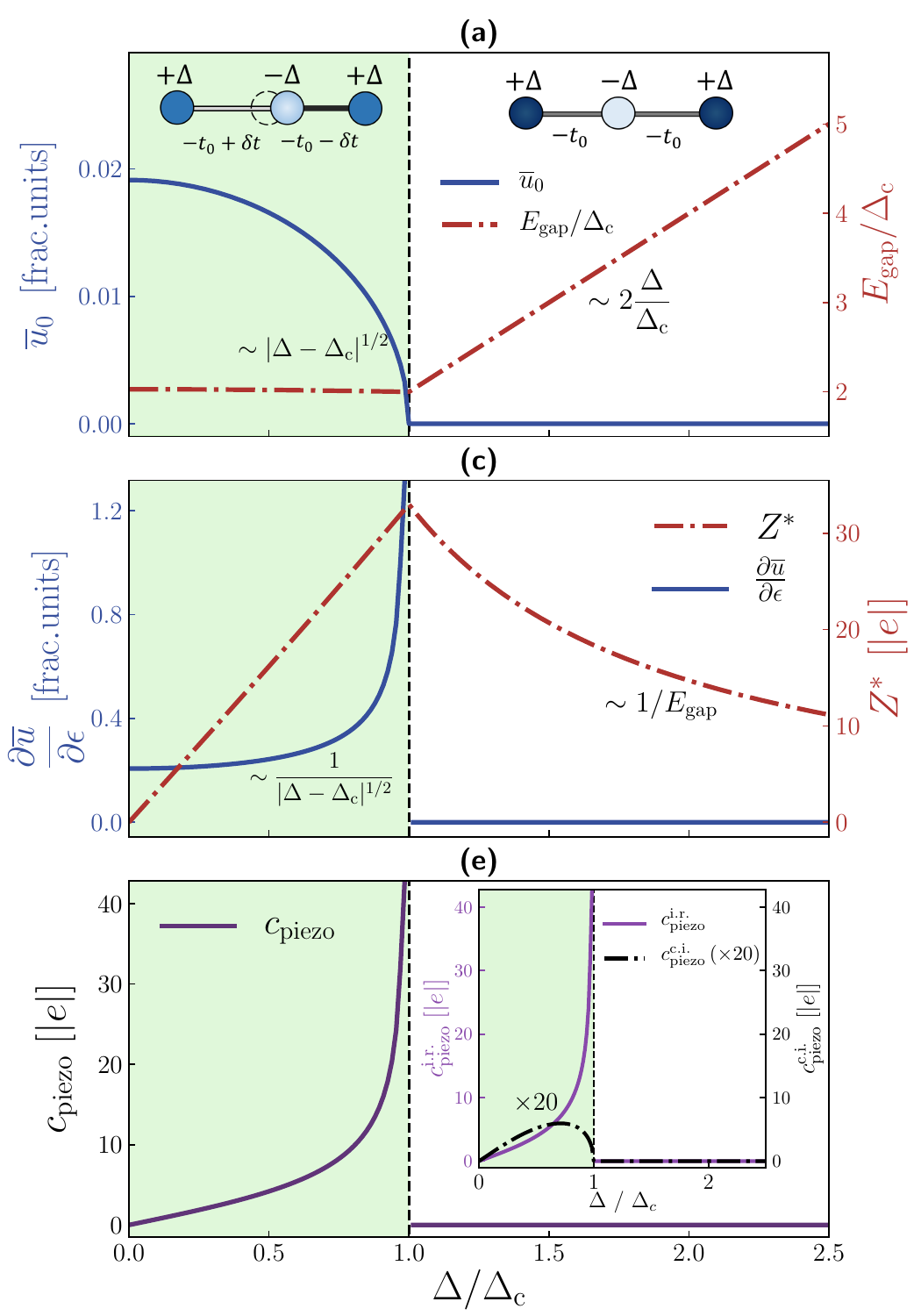}
    \includegraphics[width=0.475\textwidth]{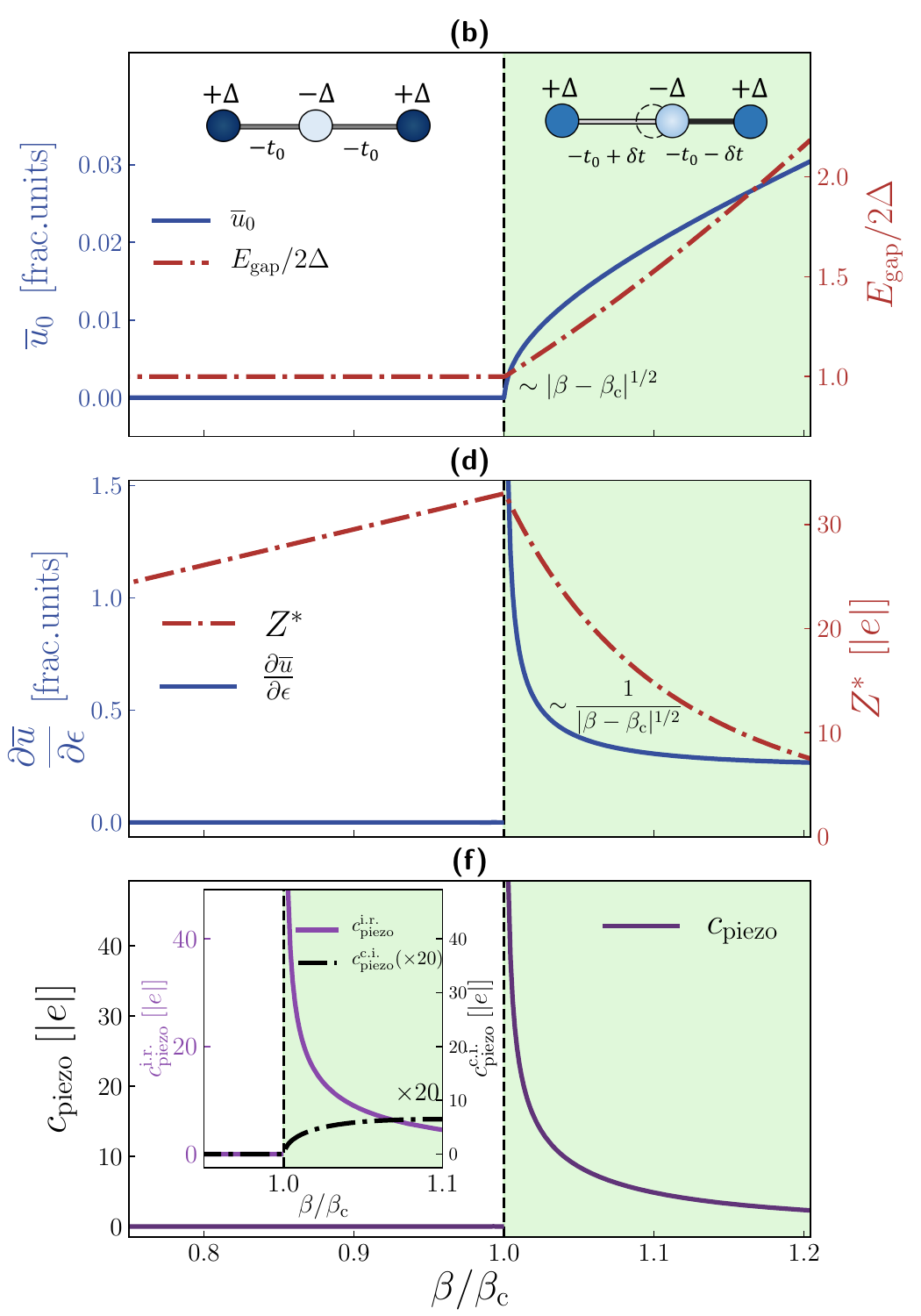}
    \end{minipage}
    \caption{\textbf{(a)} and \textbf{(b)} Structural phase transition with order parameter $\overline{u}$ as a function of the onsite energy difference $\Delta$ and of the e-ph coupling $\beta$. When $\Delta<\Delta_{\mathrm{c}}$ ($\beta>\beta_{\mathrm{c}}$) the total energy $E_{\mathrm{tot}}(u)$ has a double-well profile with two minima at $|\overline{u}|\neq0$, resulting in a distorted chain with bond-length alternation. When $\Delta>\Delta_{\mathrm{c}}$ ($\beta<\beta_{\mathrm{c}}$) the minimum of $E_{\mathrm{tot}}(u)$ is at $\overline{u}=0$ and the atoms become equidistant. The behaviour of the order parameter is typical of second order phase transitions, being $\overline{u}\propto|\Delta-\Delta_{\mathrm{c}}|^{1/2}$ ($\overline{u}\propto|\beta-\beta_{\mathrm{c}}|^{1/2}$), and the system remains insulating in both phases with gap  $E_{\mathrm{gap}} = 2\sqrt{\Delta^2 + 4 (\beta t_0\overline{u})^2}$. \textbf{(c)} and \textbf{(d)} Internal strain and effective charge across the phase transition. The inclusion of strain $\epsilon$ in the model affects the critical value $\Delta_{\mathrm{c}}(\epsilon)$ ($\beta_{\mathrm{c}}(\epsilon)$) and, as expected in a second order transition, the internal strain displays a diverging behaviour close to the critical point, $\partial\overline{u}/\partial\epsilon\propto|\Delta-\Delta_\mathrm{c}|^{-1/2}$ ($\partial\overline{u}/\partial\epsilon\propto|\beta-\beta_\mathrm{c}|^{-1/2}$). Near the critical point a huge polar response is also present, quantified by the effective charge $Z^*\propto \beta\Delta/E_{\mathrm{gap}}^2$ (see Methods). Approaching the critical point from the undimerized phase, $Z^*$ is inversely proportional to the gap $E_{\mathrm{gap}}(\overline{u}=0)\equiv\Delta$ \textbf{(c)} and linear in $\beta$ \textbf{(d)}. Since the gap is constant as a function of $\Delta$ and linear in $\beta$ when approaching the critical point from the dimerized phase, $Z^*$ displays a linear behaviour in $\Delta$ \textbf{(c)} and it is inversely proportional to $\beta$ \textbf{(d)}. \textbf{(e)} and \textbf{(f)} The piezoelectric coefficient diverges when approaching the critical point from the dimerized phase. As shown in the inset, the major contribution is due to the internal relaxation term of Equation (\ref{eq:c_model_intrel}). The topological nature of the enhancement guarantees its stability.}
    \label{fig:model_properties}
\end{figure*}

In order to have a non-trivial piezoelectric response, the chain must not have points of inversion symmetry, a requirement that is met when the equivalence between atoms is broken in a distorted chain, i.e. both $\Delta\neq0$ and $\overline{u}\neq0$. In this case, the chain becomes also ferroelectric with a net dipole moment per unit cell $P$\cite{yamauchi2014electronic}. The electromechanical response is quantified by the piezoelectric coefficient $c_{\mathrm{piezo}}$, defined as the variation of $P$ due to an applied homogeneous strain $\epsilon$, namely
\begin{align} \label{eq:c_piezo_tm}
    c_{\mathrm{piezo}} 
      = \left.\frac{\dd P(\epsilon,\overline{u}(\epsilon))}{\dd \epsilon}\right|_{\epsilon=0} = c_{\mathrm{piezo}}^{\mathrm{c.i.}} + c_{\mathrm{piezo}}^{\mathrm{i.r.}}
\end{align}
where the derivative of $P$ is decomposed in two contributions. The first one is the so-called \textit{clamped ions} term
\begin{equation} \label{eq:c_clampion_bis}
    c_{\mathrm{piezo}}^{\mathrm{c.i.}} = \left.\frac{\partial P(\epsilon,\overline{u}(\epsilon))}{\partial \epsilon}\right|_{\epsilon=0}
\end{equation}
which is obtained keeping fixed the relative position of the ions in the unit cell, i.e., for fixed internal fractional coordinate $\overline{u}_0=\overline{u}(0)$. It can be shown (see Methods) that $|c_{\mathrm{piezo}}^{\mathrm{c.i.}}| \leq |e|\beta/2\pi$. The second term of Equation (\ref{eq:c_piezo_tm}) takes into account the effect of strain on the internal coordinate $\overline{u}(\epsilon)$ and defines the \textit{internal relaxation} contribution
\begin{equation} \label{eq:c_model_intrel}
    c_{\mathrm{piezo}}^{\mathrm{i.r.}} = Z^*(\overline{u}_0)\left.\frac{\partial \overline{u}(\epsilon)}{\partial \epsilon}\right|_{\epsilon=0}
\end{equation}
where we defined the effective charge $Z^*$, namely a measure of how rigidly the electronic charge distribution follows the displacement of the nuclei, as
\begin{equation}
    \label{eq:def_Zeff_bis}
    Z^* = \frac{\partial P}{\partial u}.
\end{equation}
The inclusion of strain in the Rice-Mele model affects explicitly the critical value $\Delta_{\mathrm{c}}(\epsilon)$ of the phase transition. The behaviour of $\overline{u}(\epsilon)$ as $\Delta$ approaches $\Delta_{\mathrm{c}}(\epsilon)$ is shown in Figure \ref{fig:model_properties}a), following the expected behaviour $\overline{u}(\epsilon) \propto |\Delta - \Delta_{\mathrm{c}}(\epsilon)| ^ {1/2}$  of the order parameter of second-order phase transitions (see Supplementary Information).
It thus follows that:
\begin{equation} \label{eq:dudeps_Delta}
\left.\frac{\partial \overline{u}(\epsilon)}{\partial \epsilon}\right\vert_{\epsilon=0} \propto \frac{1}{|\Delta - \Delta_{\mathrm{c}}(0)| ^ {1/2}}.
\end{equation}
Equation (\ref{eq:dudeps_Delta}) implies that the internal relaxation term diverges as we approach the critical point $\Delta_{\mathrm{c}}(\epsilon)$ from the distorted phase, in analogy with the MPB mechanism at play in some ferroelectric oxides. Indeed, as the $\Delta$ parameter of the Rice-Mele model accounts for the composition of the system, it allows to continuously tune a morphotropic-like phase transition from the distorted phase (lower symmetry, ferroelectric) to the undistorted one (higher symmetry, paraelectric). On the other hand, at a fixed $\Delta \neq 0$ suppressing the Peierls electronic instability, the second-order phase transition can be driven by the e-ph coupling, as shown in Figure \ref{fig:model_properties}b), allowing one to define a critical value $\beta_c$ separating the undimerized and dimerized phases (see Supplementary Information). Clearly, the order parameter as a function of the variable driving the transition would still follow the expected behaviour $\overline{u}(\epsilon) \propto |\beta - \beta_c(\epsilon)|^{1/2}$, implying:
\begin{equation} 
    \label{eq:dudeps_beta}
    \left.\frac{\partial\overline{u}(\epsilon)}{\partial \epsilon}\right\vert_{\epsilon=0} \propto \frac{1}{|\beta - \beta_c(0)|^{1/2}},
\end{equation}
i.e., a diverging internal strain when approaching the critical point from the dimerized phase.

\subsection{Topological contribution to the enhancement}\label{sec:top_enhancement}
In principle, the diverging behaviour of the internal strain, Equation (\ref{eq:dudeps_Delta}) or (\ref{eq:dudeps_beta}), guarantees the existence of piezoelectric polymers with arbitrarily high response when close to a morphotropic-like phase boundary, irrespective of the prefactor, namely the effective charge $Z^*$. However, this specific enhancement is a consequence of the second order transition. Numerical evidence in linear acetylenic carbon chains\cite{romanin2021dominant} show that quantum anharmonic effects (QAE) may change the order of the structural phase transition, therefore damping the diverging behaviour of $c_{\mathrm{piezo}}^{\mathrm{i.r.}}$. A robust enhancement of the piezoelectric coefficient against QAE would depend, therefore, on the strength of the polar response embodied by $Z^*$. Measuring how the electronic charge distribution follows the displacement of the nuclei, effective charge's behaviour in this system is strictly related to the topological charge transport of the Thouless pump\cite{thouless1983quantization}. A simple geometric argument shows that indeed the effective charge $Z^*$ diverges as $1/E_{\mathrm{gap}}$ in the undimerized phase (see Methods and Supplementary Information). As first noted in Ref. \cite{onoda2004topological}, such a remarkable behaviour is a consequence of the topology of the domain of the dipole moment $P$. As $E_{\mathrm{gap}}$ goes to $0$, we get closer to the metallic point, where $P$ is not well defined, and even an infinitesimal atomic displacement causes a huge redistribution of the charge density. We contrast this result with the predicted behaviour of polar responses in 2D gapped graphene, where both piezoelectric coefficient and effective charges were found to be independent on the band-gap amplitude\cite{bistoni2019giant}. Indeed, electron-strain/lattice couplings in 2D hexagonal crystals can be described as gauge fields \cite{rostami2018npj,bistoni2019giant} whose effect is to shift the Dirac cone of an amount proportional to the coupling constants, causing the latter to be the only relevant quantities determining the strength of polar responses. In the 1D chain, instead, the e-ph interaction contributes, through dimerization, to the gap opening, thus directly affecting the Thouless-pump topological enhancement of effective charge. We further remark that the absence of a structural transition in gapped graphene causes the piezoelectric response to be mostly due to the clamped-ion contribution, the internal-relaxation one contributing roughly 25\% to the total response\cite{bistoni2019giant}.

The evolution of $Z^*$ as a function of $\Delta/\Delta_{\mathrm{c}}$ and $\beta/\beta_{\mathrm{c}}$ is shown in Figure \ref{fig:model_properties}c) and \ref{fig:model_properties}d). Even though the system always displays a finite gap preventing the metallic divergence of the effective charge, $Z^*$ reaches the giant value of $\sim \! 30\,|e|$ at the critical points. Such anomalously large effective charges can not be ascribed to the mixed covalent-ionic character of the system only: it can be shown (see Supplementary Information for the simple case of a heteropolar biatomic molecule) that tuning the bond character can lead to a finite enhancement of the effective charges, which are typically only few times the value of the nominal charge\cite{Ghosez.PhysRevB.58.6224}. Unlike the MPB-related enhancement of the internal strain, also shown in Figures \ref{fig:model_properties}c) and \ref{fig:model_properties}d), the topological behaviour is expected to be much more stable with respect to QAE, guaranteeing the enhancement of the electromechanical response. The total piezoelectric coefficient, comprising both the clamped ion and internal relaxation contributions, is shown in Figures \ref{fig:model_properties}e) and f) as a function of parameters $\Delta/\Delta_{\mathrm{c}}$ and $\beta/\beta_{\mathrm{c}}$. Insets highlight how the piezoelectric coefficient is mostly contributed by the internal-relaxation contribution, that is strongly enhanced by the combined effect of diverging internal strain $\partial\bar{u}/\partial \epsilon$ and anomalously large effective charges. We remark the importance of both mechanisms, since anomalous effective charges alone in general do not guarantee piezoelectric effects if inversion symmetry is kept, as in centrosymmetric CaTiO$_3$ and SrTiO$_3$\cite{Ghosez.PhysRevB.58.6224}, or if the internal strains are small, as in 2D hexagonal systems and gapped graphene\cite{bistoni2019giant}.

\subsection{Numerical calculations} 
\begin{figure*}[!htb]
    \begin{minipage}[c]{1.0\textwidth}
    \centering
    \includegraphics[width=0.475\textwidth]{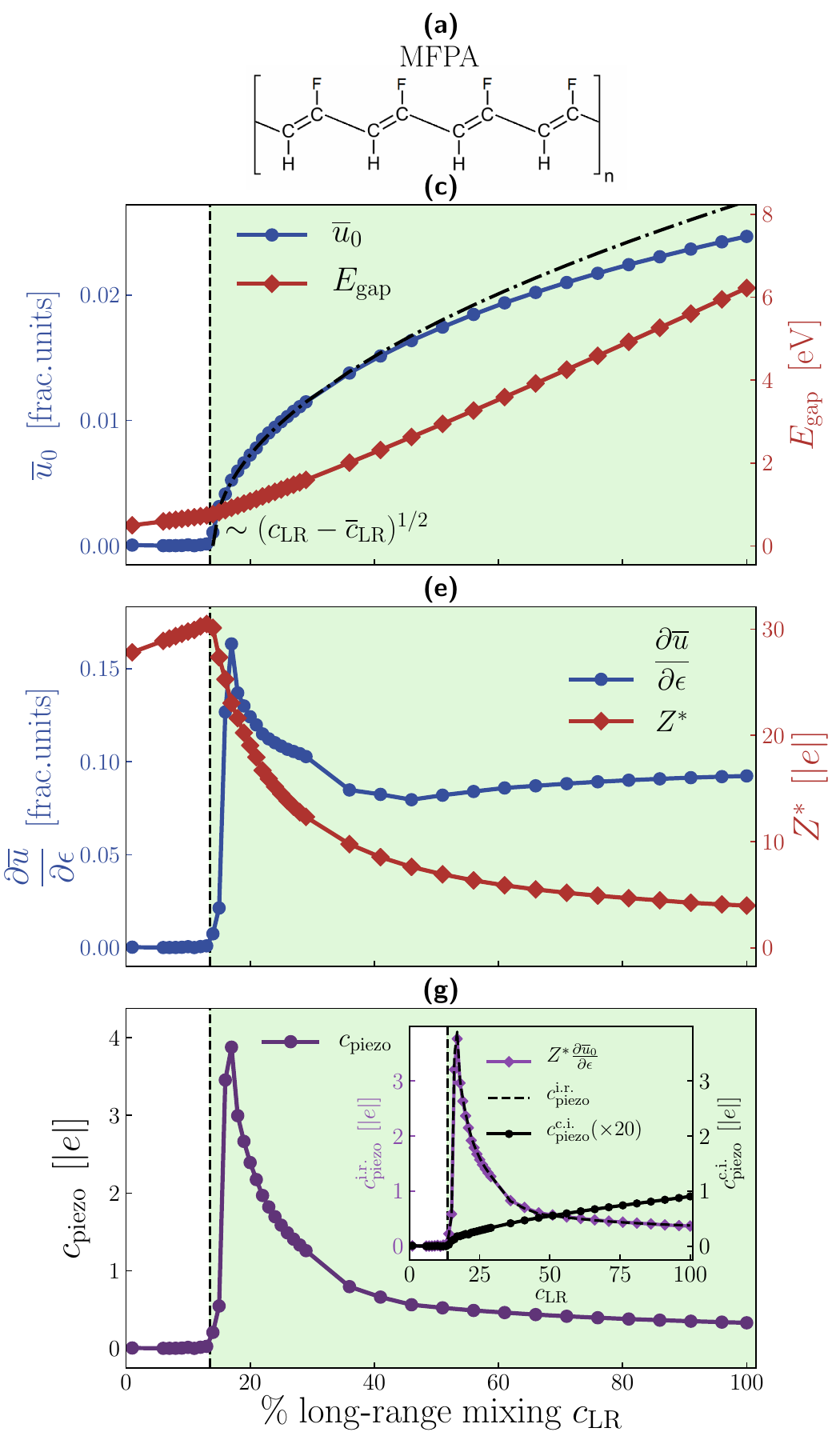}
    \includegraphics[width=0.475\textwidth]{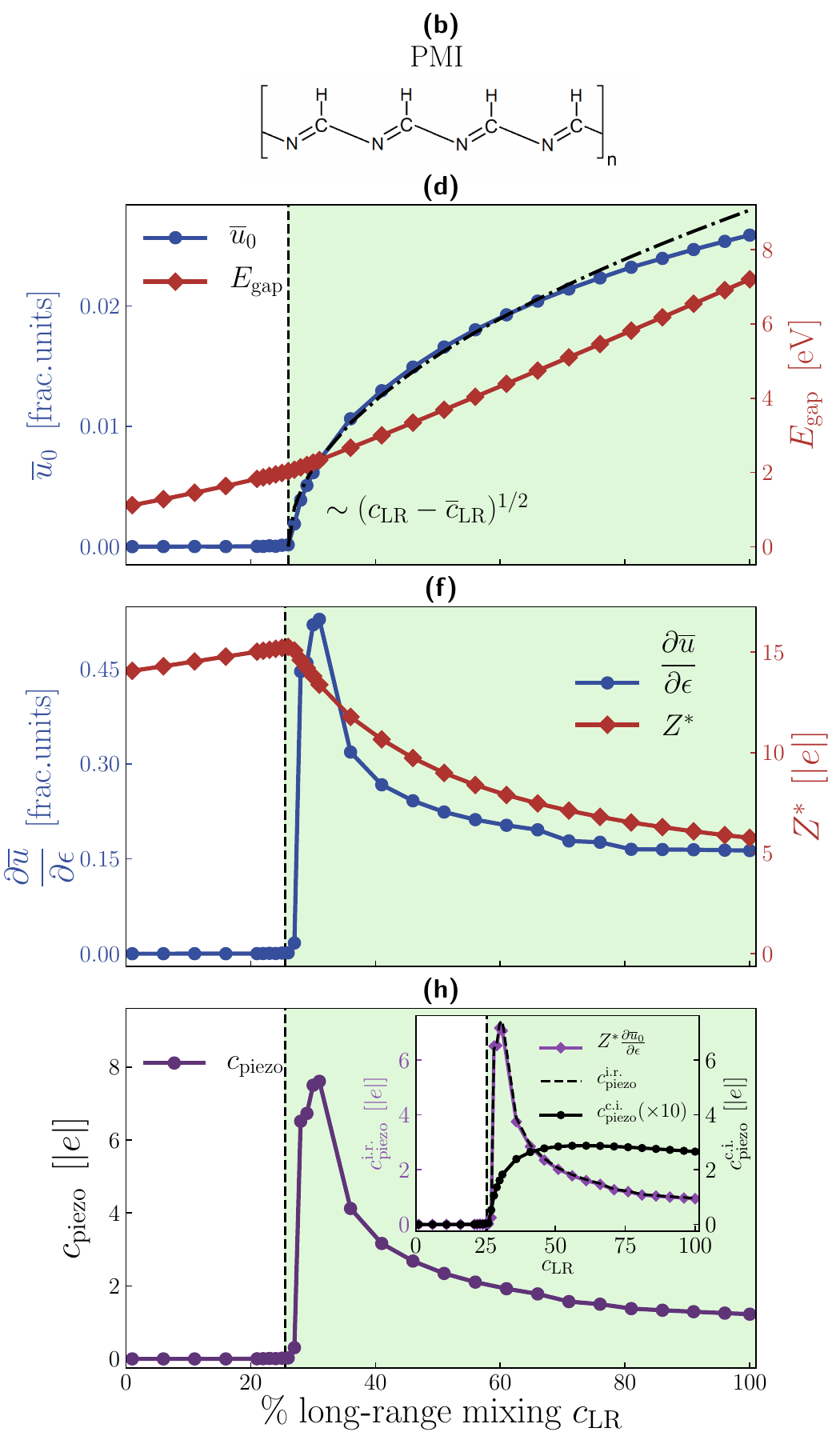}
    \end{minipage}
    \caption{\textbf{(a)} Mono-fluorinated polyacetylene (MFPA). \textbf{(b)} Polymethineimine (PMI). In \textbf{(c)} and \textbf{(d)} for MFPA and PMI, respectively, is shown the behaviour of the internal coordinate $\overline{u}_0$ for different values of the long-range mixing parameter $c_{\mathrm{LR}}$ put in the range-separated xc-functional in the DFT calculations. Consistently with the prediction of the model, we observe the behaviour $\overline{u}\simeq|c_{\mathrm{LR}}-\overline{c}_{\mathrm{LR}}|^{1/2}$, with $\overline{c}_{\mathrm{LR}}^{\text{MFPA}}\simeq14\%$ and $\overline{c}_{\mathrm{LR}}^{\text{PMI}}\simeq26\%$. In \textbf{(e)} and \textbf{(f)}, for MFPA and PMI respectively, the behaviour of the $\partial\overline{u}_0/\partial \epsilon$ is reported, along with the values of the effective charge $Z^*$. For each polymer, we chose $Z^* = Z^*_{C,xx}$ (for MFPA the C bound to the F). In agreement with the model, on the one hand we observe a further hint of the morphotropic-like nature of the transition while on the other hand the huge values of $Z^*$ stands out, in particular in the region near the critical points. In \textbf{(g)} and \textbf{(f)} is shown how the piezoelectric coefficient is greatly enhanced when reaching the critical points from the less symmetric phase. The comparison between $c_{\mathrm{piezo}}^{\mathrm{c.i.}}$ and $c_{\mathrm{piezo}}^{\mathrm{i.r.}}$, in the insets, highlights the internal-relaxation origin of the enhancement. Furthermore, the comparison with the results obtained putting the values of \textbf{(e)} and \textbf{(f)} in Equation (\ref{eq:c_model_intrel}) shows that the model very well describes the nature of the enhancement.}
    \label{fig:abinitio_simulation}
\end{figure*}
We performed ab initio calculations in the framework of Density Functional Theory (DFT) to validate our model predictions, choosing two conjugate polymers representative of the broad class of SPA. One is monofluorinated polyacetylene (MFPA) made by the repetition of the unit CH-CF, obtained by substituting one hydrogen atom of the C$_2$H$_2$ unit of PA with fluorine. The other is polymethineimine (PMI), obtained substituting a CH pair with a nitrogen atom to obtain the unit N-CH. For simplicity, we considered the all-trans structures shown in Figure \ref{fig:abinitio_simulation}a) and \ref{fig:abinitio_simulation}b), whose fundamental physical properties are captured by the Rice-Mele model. Even though controlling the fraction of substituted atoms may in principle induce a morphotropic-like transition, this approach poses many challenges both from the computational and experimental side: to our specific purposes, it wouldn't allow to study the phase transition and the associated predicted enhancement of piezoelectric effect by varying with continuity an external parameter (as $\Delta$ in the model). As discussed in sections \ref{sec:MPB_enhancement} and \ref{sec:top_enhancement}, the internal-strain and the effective-charge enhancements may be also induced by tuning the parameter $\beta$. To achieve this computational task, we take advantage of the effect of screened Coulomb vertex corrections to the dressing of the e-ph coupling\cite{LuoieGWPT_PRL2019}, leading to an enhancement of e-ph itself especially strong in low-dimensional materials and when phonons at zone boundary are involved \cite{Basko.PhysRevB.77.041409,Attaccalite_prb2008,PhysRevLett.130.256901,Pamuk.PhysRevB.2016}. Such screened-Coulomb-mediated e-ph enhancement can be captured by hybrid functionals incorporating a fraction of the exact exchange \cite{Attaccalite_prb2008,cohen.prb2010}. Its inclusion has been proven essential for describing the bond-length alternation of trans-polyacetylene \cite{JACQUEMIN2005376,ferretti2012ab} and related 1D polymers \cite{romanin2021dominant, Cudazzo_prb2022}, whose BLA is typically underestimated by standard local-density or generalized-gradient approximations, indirectly pointing to an enhancement of the e-ph coupling due to electron-electron interaction. Range-separated hybrid (RSH) functionals represent an ideal choice for our purpose, as they have been designed to better account for the screened Coulomb vertex corrections. The latter can be effectively tuned by acting on the long-range (LR) mixing parameter $c_{\mathrm{LR}}$ that accounts for the fraction of LR exact exchange in RSH functionals, thus providing a computational knob to continuously vary $\beta$. We further remark that the strength of e-ph enhancement due to screened Coulomb effects can be ideally controlled by modifying the screening itself, as proposed for doped graphene \cite{attaccalite2_nanoletters2010}. Since the optimal mixing parameter $c_{\mathrm{LR}}$ is inversely proportional to the scalar dielectric constant of the environment in order to enforce the correct asymptotic potential \cite{govonigalli_PhysRevX.6.041002_2016, kronik.advmat2018,refaely.prb2013,luftner.prb2014,manna.jctc2018}, we speculate that controlling the dielectric environment may represent a viable strategy, alternative and complementary to controlling the fraction of substituted atoms, for tuning and optimising the piezoelectric response of conjugated polymers.

Motivated by these reasons, we performed structural optimization of both MFPA and PMI for different values of the LR mixing parameter $c_{\mathrm{LR}}$. For consistency with the model, we considered the coordinates of the C and N atoms along the principal axis of the chain, which we take as the $x$-axis, to compute the internal coordinate. More details on the effect of $c_{\mathrm{LR}}$ on polymers' structures are provided in the Supplementary Information. The evolution of $\overline{u}_0$ displayed in Figure \ref{fig:abinitio_simulation}c) and \ref{fig:abinitio_simulation}d), clearly hints to the presence of a second order phase transition triggered by $c_{\mathrm{LR}}$ for both polymers: the dimerized phase is suppressed by lowering the fraction of mixing, the order parameter showing the expected behaviour as it approaches the second-order phase-transition critical point, in excellent qualitative agreement with model results shown in Figure \ref{fig:model_properties}b), d) and f) and confirming a posteriori the direct proportionality between $c_{\mathrm{LR}}$ and $\beta$. The second-order character of the phase transition is further confirmed by the softening of the corresponding optical phonon found in the higher-symmetry phase when increasing $c_{\mathrm{LR}}$ (Figure S6), signalling the onset of a dynamical instability of the undimerized structure.

In Figure \ref{fig:abinitio_simulation}e) and \ref{fig:abinitio_simulation}f) the behaviour of $Z^*$ and of $\partial \overline{u}/\partial \epsilon$ calculated from first principles is shown. For MFPA we took $Z^* = Z^*_{C_F,xx}$ where $C_F$ is the carbon atom bound to the fluorine, while for PMI $Z^* = Z^*_{C,xx}$. The full tensors of the effective charges of all the atoms are reported in the Supplementary Information. We highlight the qualitative agreement with the prediction of the model, in particular the huge enhancement of the effective charges around the critical point $\overline{c}_{\mathrm{LR}}$, reaching the strongly anomalous values of $\sim \!\! 30\,|e|$ and $\sim \!\! 15\,|e|$ in correspondence of $\overline{c}_{\mathrm{LR}}$ for MFPA and PMI, respectively. The covalent character of bonds along the chain prevents a precise definition of the nominal reference value for C, that can be however assumed to be of the order of $1\vert e \vert$, as the nominal ionic charges for H and F are respectively $+1\vert e \vert$ and $-1\vert e \vert$. Effective charges of carbon in both considered chains are strongly anomalous for all considered long-range mixing parameters, displaying values between $5\vert e \vert$ and $30\vert e \vert$ even for band gaps exceeding 6 eV. These anomalous values exceed even those reported in oxide ferroelectrics, where effective charges are typically two or three times larger than nominal reference values\cite{Ghosez.PhysRevB.58.6224}. Figure \ref{fig:abinitio_simulation}g) and \ref{fig:abinitio_simulation}h) display the behaviour of the piezoelectric coefficients computed ab initio taking into account also the effects of transverse displacements. In the insets, the different contributions $c_{\mathrm{piezo}}^{\mathrm{i.r.}}$ and $c_{\mathrm{piezo}}^{\mathrm{c.i.}}$ are compared, highlighting the internal-relaxation origin of the enhancement. Using the the ab initio values of $Z^*$ and $\partial \overline{u}/\partial \epsilon$ of Figure \ref{fig:abinitio_simulation}e) and \ref{fig:abinitio_simulation}f), we compare the values of $c_{\mathrm{piezo}}^{\mathrm{i.r.}}$ computed without approximations with those obtained using Equation (\ref{eq:c_model_intrel}) of the model. The agreement between the two approaches is both qualitatively and quantitatively excellent, notwithstanding the simplifying description provided by the Rice-Mele model, that neglects structural details specific of the two considered polymers as well as transverse displacements. We highlight that despite the behaviour $\partial \overline{u}/\partial \epsilon\propto|c_{\mathrm{LR}} - \overline{c}_{\mathrm{LR}}|^{-1/2}$, the main contribution to the piezoelectric coefficient is given by the effective charges. The large values attained in a finite range around the second-order critical point and their ultimately topological origin suggest that the piezoelectric effect is robust against quantum and anharmonic effects that may change the order of the phase transition, as predicted in carbyne\cite{romanin2021dominant}.

\begin{figure}[!htb]
    \begin{minipage}[c]{1.0\linewidth}
    \centering
    \includegraphics[width=1.0\textwidth]{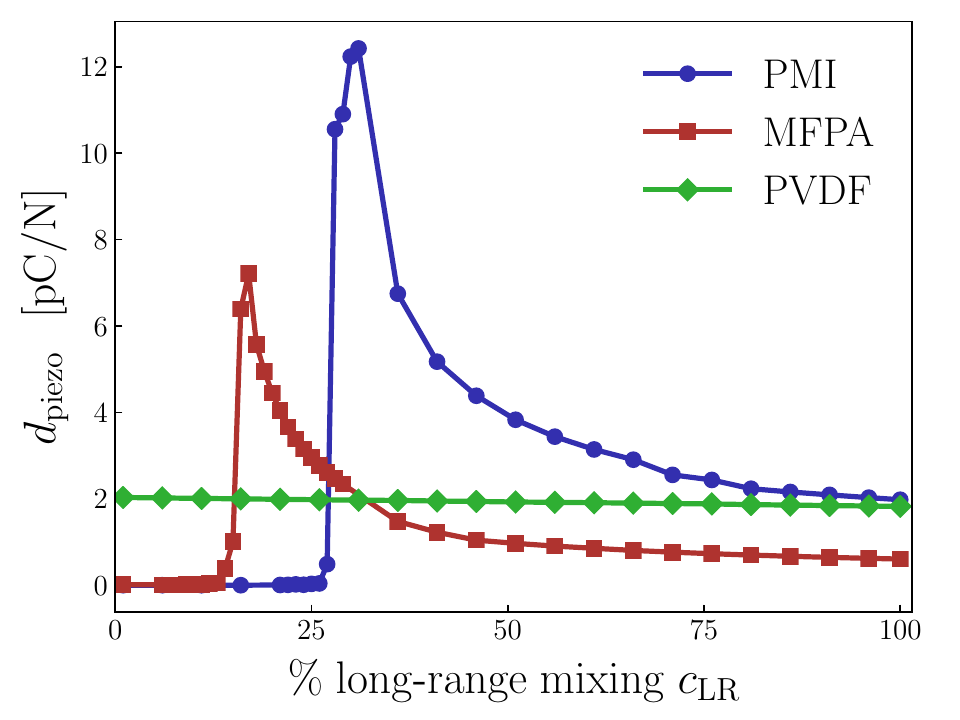}
    \end{minipage}
    \caption{Comparison of the converse piezoelectric coefficients of MFPA and PMI with respect to PVDF, the current best and most widely used organic piezoelectric. In principle, the mechanism of the morphotropic-topological enhancement allows to outperform the state-of-the-art.}
    \label{fig:piezo_comparison}
\end{figure}

\section*{Discussion}
Finally, we compare the results for the piezoelectric coefficients of MFPA and PMI with those of the best and most widely used piezoelectric PVDF polymer, made by the repetition of the unit CF$_2$-CH$_2$. The Rice-Mele model fails to capture its main properties, this polymer being not conjugated. Its piezoelectricity indeed derives from the presence of a net dipole moment transverse to the chain, whereas the electromechanical response predicted in conjugated polymers is longitudinal to the chain and ultimately due to the topological-morphotropic enhancement. To have a consistent comparison with available experimental data for PVDF, we computed ab initio the \textit{converse} piezoelectric coefficient $d_{\mathrm{piezo}}$, which measures the response with respect to an external stress, rather than a strain (see Methods). The results for the converse piezoelectric coefficients of PVDF are compared with those of MFPA and PMI in {Figure \ref{fig:piezo_comparison}} and are consistent with the values computed in Ref. \cite{nakhmanson2004ab}. Even though the calculated $d_{\mathrm{piezo}}$ is smaller than reported experimental values, a direct comparison to experiments is hardly drawn because, e.g., of the polymorphic character or low crystallinity of experimental samples, as noticed also in Ref. \cite{nakhmanson2004ab}. We remark that piezoelectric response in PVDF is found to be independent on the fraction of exact exchange, confirming the utterly different nature of the electromechanical response in such non-conjugated polymer. We finally mention that mildly anomalous effective charges have been also reported for PVDF\cite{RAMER200510431}, the carbon effective charge however not exceeding 1.5$|e|$, consistently with our results provided in Supplementary Information. On the other hand, both MFPA and PMI display a rather large range of values that are larger than calculated $d_{\mathrm{piezo}}$ of PVDF, with up to a six-fold enhancement for PMI close to the dimerization point. The robustness of the enhancement mechanism is confirmed also by the comparison with the piezoelectric coefficients calculated in packed MFPA chains, as discussed in Supplementary Information. Even though the selected prototypical SPAs may not be the most efficient ones for practical realization and engineering of their functional properties, the comparison with first-principles estimate of one of the best available piezoelectric polymer alongside the general validity of the proposed model and consequent robustness of its electromechanical response put forward the broad class of $\pi$-conjugated polymers as a promising field for organic piezoelectrics with enhanced functionalities. Additionally, its inverse proportionality to the band gap provides a possible material-design principle for driving the quest of organic polymers with enhanced piezoelectric response. 

\section*{Methods}
\subsection*{Details on the model}
The Hamiltonian of the chain is the sum of two terms $ H_{\mathrm{tot}} = H_{\mathrm{L}} + H_{\mathrm{e}}$. The lattice contribution $H_{\mathrm{L}}$ accounts for the displacement of the atoms with respect to their position in a uniformly spaced chain: 
\begin{equation} \label{eq:H_ionic}
    H_{\mathrm{L}} = \sum_{i}\frac{p_i^2}{2m_i} + \frac{1}{2}\mathrm{K}\sum_{i}(\delta r_{i+1} - \delta r_i)^2
\end{equation}
where $m_i$ and $\delta r_i$ are mass and displacement of atom $i$, $p_i$ its momentum, and $\mathrm{K}$ is a spring constant term. In a nearest neighbours tight-binding approximation, the electronic contribution $H_{\mathrm{e}}$ reads
\begin{equation} \label{eq:H_electronic}
    H_{\mathrm{e}} = 2\sum_{i}\left[\Delta(-1)^{i}c_{i}^{\dagger}c^{\phantom{\dagger}}_{i} - \left( t_{i,i+1} c_{i}^\dagger c^{\phantom{\dagger}}_{i+1} + \mathrm{h.c.} \right) \right],
\end{equation}
where the factor $2$ accounts for the spin degeneracy, $c_i^{\dagger}$/$c_i$ are creation/annihilation operators for electrons, $t_{i,i+1}$ is a hopping energy while $\Delta>0$ accounts for the onsite energy difference between neighbours. The adimensional order parameter $u(\epsilon)$ measures the relative displacement between neighbours and is defined as
\begin{equation}
    \label{eq:def_u_eps}
    u(\epsilon) = \frac{\delta r_{i+1}(\epsilon) - \delta r_{i}(\epsilon) }{a(\epsilon)/2}.
\end{equation}
The total energy of the system $E_{\mathrm{tot}}(u)$ as a function of the order parameter $u$ has two contributions, namely
\begin{equation}
    \label{eq:def_Etot_vs_u}
    E_{\mathrm{tot}}(u) = E_{\mathrm{L}}(u) + E_{\mathrm{e}}(u).
\end{equation}
The first term $E_{\mathrm{L}}(u)$ accounts for lattice distortion and is quadratic in $u$, whereas the electronic contribution $E_{\mathrm{e}}(u)$ is linear in $u$. For a given set of material-dependent parameter $t_0$, $\mathrm{K}$, $\Delta$, $\beta$ and the strain $\epsilon$, we find the optimal displacement $\overline{u}$ -- defined as the one which minimises the total energy -- as shown in detail in the Supplementary Information.

We computed the dipole moment per unit cell $P$ using the Berry phase approach\cite{king1993theory,resta2000manifestations}. In the limit $E_{\mathrm{gap}} \ll t(\epsilon)$ it holds (see Supplementary Information)
\begin{align} \label{eq:polarisation_model}
    P(\epsilon,\overline{u}(\epsilon)) = -\frac{|e|}{\pi}\theta(\epsilon,\overline{u}(\epsilon))
\end{align}
with  $\tan\theta(\epsilon,\overline{u}(\epsilon)) = \Delta/2\delta t(\epsilon,\overline{u}(\epsilon))$, where we emphasize that the dependence on the strain enters both directly and through $\overline{u}(\epsilon)$. Using Equation (\ref{eq:def_dt(eps)}) and (\ref{eq:polarisation_model}) we have
\begin{equation} \label{eq:c_clampion}
    c_{\mathrm{piezo}}^{\mathrm{c.i.}} = -\frac{|e|}{2\pi}\beta\sin 2\theta(0,\overline{u}_0)
\end{equation}
where we notice that $|c_{\mathrm{piezo}}^{\mathrm{c.i.}}| \leq |e|\beta/2\pi$, the maximum achievable value being directly proportional to the e-ph coupling constant $\beta$.

Plugging Equation (\ref{eq:def_dt(eps)}) in Equation (\ref{eq:polarisation_model}), we obtain an explicit expression for the effective charge in both the dimerized and undimerized phases:
\begin{equation} \label{eq:def_Zeff}
    Z^*(\overline{u}_0) = \frac{|e|}{\pi}4\beta t_0\frac{\sin\theta(0,\overline{u}_0)}{E_{\mathrm{gap}}(0,\overline{u}_0)} 
\end{equation}
where $\sin\theta=2\Delta/E_{\mathrm{gap}}$. In the undimerized phase $\overline{u}=0$ implies $\sin\theta=1$, hence $Z^*\sim \beta/E_{\mathrm{gap}}\sim\beta/\Delta$. The diverging behaviour can be further understood using a simple geometric argument that highlights its topological origin. From Equation (\ref{eq:polarisation_model}), polarization is proportional to an angle $\theta$ spanning the parametric ($\Delta,\delta t$)-space along a circumference with radius $E_{\mathrm{gap}}$. As $\delta t\propto u$, it follows that in the undimerized phase $E_{\mathrm{gap}} d\theta\propto du$, hence $Z^*=\partial P/\partial u \propto\partial \theta/\partial u\propto 1/E_{\mathrm{gap}}$ (more details in the Supplementary Information).

The parameters used to produce the results in {Figure \ref{fig:model_properties}} were obtained fitting the model with the PBE0\cite{perdew1996rationale} total energy profile of carbyne as a function of the relative displacement of the two carbon atoms of its unit cell. In particular, with the value $a_0=2.534$ \AA~ obtained through a cell-relaxation procedure, the fit yields $t_0=2.239$ eV, $\beta=3.906$ and $\mathrm{K}=127.9$ eV/\AA$^2$, whereas for carbyne it holds $\Delta=0$. The behaviour with respect to $\beta/\beta_c$ was obtained fixing a finite $\Delta=1.05\Delta_c$. We highlight that each carbon atom of carbyne contributes with two electrons to the $\pi$-orbital, so it is necessary to put an additional factor of $2$ in front of Equation (\ref{eq:H_electronic}) to account for this degeneracy. 

\subsection*{Computational details}
All DFT calculations were performed using CRYSTAL code\cite{dovesi2014crystal14,dovesi2018quantum}, which employs a basis of local Gaussian-type functions. This approach allows for the simulation of truly isolated systems, as the 1D polymers addressed in our work, using the hybrid functionals, proven to be essential for accurately reproducing the physics of 1D chains, \cite{romanin2021dominant,ferretti2012ab}. The Gaussian-type basis set significantly reduces the computational cost of evaluating real-space integrals and, hence, it allows to drastically reduce the computational cost when compared, e.g., with state-of-the-art plane-waves based DFT codes. We used a triple-$\zeta$-polarised Gaussian-type basis\cite{vilela2019bsse} with real space integration tolerances of 10-10-10-15-30 and an energy tolerance of $10^{-10}$ Ha for the total energy convergence. We customised a range-separated LC-$\omega$PBE hybrid exchange-correlation functional\cite{weintraub2009long} varying the value of the long-range (LR) mixing parameter $c_{\mathrm{LR}}$ which enters in the definition of the LR part of the functional, namely
\begin{equation}
    E_{\mathrm{xc}}^{\mathrm{LC-}\omega\mathrm{PBE}} = E_{\mathrm{xc}}^{\mathrm{PBE}} + c_{\mathrm{LR}}\left(E_{\mathrm{x}}^{\mathrm{LR,HF}} - E_{\mathrm{x}}^{\mathrm{LR,PBE}}\right).
\end{equation}
When $c_{\mathrm{LR}}=0$ the PBE functional is recovered while if $c_{\mathrm{LR}}=1$ we have pure Hartree-Fock (HF) exchange. The long-range terms in round brackets depend on the range-separation parameter $\omega$ that enters in the decomposition of the Coulomb operator $1/r$ as
\begin{equation} \label{eq:RSH_crystal}
        \frac{1}{r} = \frac{1-\mathrm{erf}(\omega r)}{r} + \frac{\mathrm{erf}(\omega r)}{r}
\end{equation}
where $\mathrm{erf}(\cdot)$ is the error function; the first and second term in the right-hand side of Eq. \ref{eq:RSH_crystal} account for the short- and long-range part of the Coulomb operator, respectively. All the presented values were obtained with $\omega = 0.4$ a$_0^{-1}$. For each $c_{\mathrm{LR}}$, a geometric optimisation was performed and all quantities were computed on the equilibrium configurations (see Supplementary Information). The derivatives $\partial \overline{u}/\partial \epsilon$ of the order parameter with respect to the strain were computed with finite differences, performing a fixed-cell optimisation for each strained configuration with cell length $a(\epsilon)=a_0(1\pm\epsilon)$ and $\epsilon=0.01$, while the effective charges were computed as finite differences of polarization. Values of polarization along the chain (parallel to the $x$-axis) were computed using the Berry phase approach, whereas components transverse to the isolated chain were computed in real space.\cite{zicovich2004calculation,pascale2004calculation}. The piezoelectric coefficients $c_{\mathrm{piezo}}$ and $d_{\mathrm{piezo}}$ of Figure \ref{fig:abinitio_simulation}g), \ref{fig:abinitio_simulation}h) and \ref{fig:piezo_comparison}, as well as the values of $c_{\mathrm{piezo}}^{\mathrm{c.i.}}$ and of $c_{\mathrm{piezo}}^{\mathrm{i.r.}}$, were computed using the Berry-phase approach\cite{vanderbilt2000berry} as implemented in the code\cite{erba2013piezoelectricity,erba2016internal}, which accounts also for transverse displacements. The \textit{converse} piezoelectric coefficient $d_{\mathrm{piezo}}$ is linearly related to $c_{\mathrm{piezo}}$ through the elastic constants tensor $\mathbb{C}$, namely $c_{\mathrm{piezo}} = d_{\mathrm{piezo}}\mathbb{C}$. As far as 1D systems are concerned, only a single scalar elastic constant is required, and it can be evaluated as the second derivative of the energy with respect to the strain, i.e. $\mathbb{C}=\partial^2 E/\partial \epsilon^2$. 

\medskip
\section*{Acknowledgements} \par
The authors acknowledge financial support from the European Union under ERC-SYN MORE-TEM, No. 951215, and from the Italian MIUR through PRIN-2017 project, Grant No. 2017Z8TS5B.
We also acknowledge CINECA awards under ISCRA initiative Grant No. HP10CCJFWR and HP10C7XPLJ for the availability of high performance computing resources and support. Views and opinions expressed are however those of the author(s) only and do not necessarily reflect those of the European Union or the European Research Council. Neither the European Union nor the granting authority can be held responsible for them.

\section*{Competing interests}
The Authors declare no Competing Financial or Non-Financial Interests.

\section*{Data availability}
The original data for each figure are available from the corresponding author upon request.

\section*{Code availability}
All numerical calculations were performed using the CRYSTAL code. Detailed information related to the license and user guide are available here \texttt{https://www.crystal.unito.it/}.

\section*{Authors contribution}
\textbf{S.P.V.}: theory development and implementation, numerical calculations and analysis, original draft writing.
\textbf{M.C.}: numerical calculations and validation.
\textbf{P.B.}: conceptualization, validation, supervision, original draft writing.
\textbf{F.M.}: methodology, conceptualization, validation, supervision, and project co-ordination.
All co-authors contributed to the review and editing of the manuscript.

\medskip

\bibliographystyle{unsrt}
\bibliography{bibliography}

\appendix
\onecolumngrid

\counterwithin{figure}{section}
\renewcommand{\thefigure}{S\arabic{figure}}
\setcounter{figure}{0}

\renewcommand{\thetable}{S\arabic{table}}

\setcounter{equation}{0}
\renewcommand{\theequation}{S\arabic{equation}}

\renewcommand{\thesubsection}{S\arabic{subsection}}

\begin{center}
\textcolor{white}{}\\
\vspace{1cm}
\large\textbf{Supplementary Information}
\end{center}
\vspace{0.cm}

\section*{Inclusion of strain in the Rice-Mele model}
In this section we provide a detailed description of the Rice-Mele model\cite{rice1982elementary} and the proposed extension that enables the description of strain effects. We consider an infinitely long one-dimensional linear chain made by the repetition of a unit cell, of length $a_0$, containing two atoms: one of type $A$ and the other of type $B$. We recall that the only possible strains in 1D are contractions or dilatations of the unit cell. Defining the adimensional parameter $\epsilon$, the effect of strain on the unit cell length $a_0$ is 
\begin{equation}
    \label{eq:def_a_epsilon}
    a(\epsilon)=a_0(1+\epsilon).    
\end{equation}
Electronic properties of the system are described in a nearest-neighbour tight-binding approximation. We assume one electronic orbital per atom, e.g., the $p$-orbital of carbon atoms, and we adopt the notation $\ket{\alpha,R}$ to indicate that the orbital of atom $\alpha=A,B$ is located at $r_{\alpha}+R$, $R=na$ ($n\in \mathbb{Z}$) being the position of the atom's cell along the chain. Without loss of generality, we take $r_A=-a/4 + \delta r_A$ and $r_B=+a/4 + \delta r_B$, $\delta r_{\alpha}$ being the displacement of atom $\alpha$. For simplicity, we consider only longitudinal displacements, parallel to the linear-chain direction. The basis set of the orbitals $\{\ket{\alpha,R}\}$ is orthonormal and it holds:
\begin{equation}
    \Braket{\alpha,R | \alpha',R'} = \delta_{\alpha,\alpha'} \delta_{R,R'}.
\end{equation}
We define the onsite energy terms of the electronic Hamiltonian $H_{\mathrm{e}}$ as
\begin{equation}
    \label{eq:def_Delta}
    \Braket{A,R|H_{\mathrm{e}}|A,R'} = -\Braket{B,R|H_{\mathrm{e}}|B,R'} = -\Delta \delta_{R, R'}, \; \mathrm{with} \;\; \Delta \geq 0.
\end{equation}
We take into account also the energetic contribution due to the overlap between an atom's orbital and the orbitals of its left and right nearest neighbours. In general, the overlap energy term between two electronic orbitals is a function of the distance $r$ between the atoms and is referred to as the hopping energy $-t(r)$, $t>0$. For convenience, we indicate with $r_1$ the distance between atoms in the same cell and with $r_2$ the distance between neighbouring atoms in adjacent cells, namely: 
\begin{align}
    \label{eq:def_r1}
    r_1 &\equiv (r_B + R) - (r_A + R) = \frac{a}{2} + \delta r_B - \delta r_A,
    \\
    \label{eq:def_r2}
    r_2 &\equiv (r_A + R) - (r_B + R - a) = \frac{a}{2} + \delta r_A - \delta r_B
\end{align}
and it holds $r_1+r_2=a$. We can now define the hopping energy between orbitals of atoms in the same cell $-t_1\equiv-t(r_1)$ and the hopping energy between atoms in adjacent cells $-t_2\equiv-t(r_2)$. With the same notation of {Equation (\ref{eq:def_Delta})} we write:
\begin{equation}
    \label{eq:def_t1_t2}
    \Braket{A,R|H_{\mathrm{e}}|B,R'} = \Braket{B,R'|H_{\mathrm{e}}|A,R}^* = -t_1\delta_{R,R'} - t_2\delta_{R-a,R'}.
\end{equation}
As we are interested in the effects of strain $\epsilon$ and of atoms' displacement, we define, using {Equation (\ref{eq:def_r1})} and {(\ref{eq:def_r2})}, an adimensional fractional coordinate $u(\epsilon)$:
\begin{equation}
    \label{eq:def_u}
    u(\epsilon) = \frac{r_1 - r_2}{a(\epsilon)} = \frac{\delta r_B - \delta r_A}{a(\epsilon)/2}.
\end{equation}
This term quantifies deviations from the equally spaced chain, allowing us to express bond lengths with the following compact expression:
\begin{equation}
    \label{eq:def_ri}
    r_i = \frac{a(\epsilon)}{2} \left[ 1 + (-1)^{i+1}u(\epsilon) \right], \; i=1,2.
\end{equation}
With the above definitions, at linear order in atoms' displacement we have 
\begin{equation}
    \label{eq:def_t_i}
    t_i = t(r_i) \simeq t\left(\frac{a(\epsilon)}{2}\right) + \left.\frac{\dd t}{\dd r}\right|_{\frac{a(\epsilon)}{2}} \cdot \left(r_i - \frac{a(\epsilon)}{2}\right),
\end{equation}
which allow us to define the two terms
\begin{align}
    \label{eq:def_t_epsilon_bis}
    t(\epsilon) &= \frac{t_1 + t_2}{2}, \\
    \label{eq:def_deltat_epsilon}
    \delta t(\epsilon) &= \frac{t_1 - t_2}{2}.
\end{align}
The former term $t(\epsilon)$ quantifies the effect of strain on the hopping energy between equidistant atoms while the latter term $\delta t(\epsilon)$ describes the variation with respect to $t(\epsilon)$ caused by atoms’ relative displacement. In absence of strain ($\epsilon=0$) we recover the quantities defined in the Rice-Mele model, whereas at linear order in $\epsilon$ it holds:
\begin{align}
    t\left(\epsilon\right) &= t\left(\frac{a_0}{2}(1+\epsilon)\right) \simeq t\left( \frac{a_0}{2} \right) + \left.\frac{\dd t}{\dd r} \right|_{\frac{a_0}{2}} \cdot \left.\frac{\dd a(\epsilon)}{\dd \epsilon}\right|_{\epsilon=0} \cdot \frac{\epsilon}{2} = \\
    \label{eq:def_t_epsilon}
    & = t_0(1-\beta\epsilon)
\end{align}
where $t_0 \equiv t(a_0/2)$ and we defined the adimensional parameter $\beta>0$ as 
\begin{equation}
    \label{eq:def_beta}
    \beta \equiv - \frac{a_0}{2t_0}\left.\frac{\dd t}{\dd r}\right|_{\frac{a_0}{2}}.
\end{equation}
This term quantifies the variation of the hopping energy due to a variation of the distance between the atoms and as such it acts as an electron-phonon coupling term. Analogously, we obtain 
\begin{equation}
    \label{eq:def_deltat_epsilon_bis}
    \delta t(\epsilon) = - t_0 \beta' (1+\epsilon) u(\epsilon),
\end{equation}
where we define another adimensional e-ph parameter $\beta'>0$ as 
\begin{equation}
    \label{eq:def_beta_bis}
    \beta' = - \frac{a_0}{2t_0}\left.\frac{\dd t}{\dd r}\right|_{\frac{a(\epsilon)}{2}} .
\end{equation}
Even though the two e-ph parameters $\beta'$, $\beta$ can differ at finite values of the strain, at the lowest order one can safely assume that they coincide and consider $\beta'=\beta$. A schematic representation of the model is shown as an inset in {Figure \ref{fig:Etot_vs_u}}.

\section*{Structural phase transition}
As we are interested in the structural properties at $T= 0~K$, we study the total energy per unit cell $E_{\mathrm{tot}}(u) = E_{\mathrm{L}}(u) + E_{\mathrm{e}}(u)$, where $E_{\mathrm{L}}(u)$ and $E_{\mathrm{e}}(u)$ are the lattice and electronic contribution, respectively. In particular, we aim to characterise the behaviour of the optimal displacement $\overline{u}$, defined as the one which minimises $E_{\mathrm{tot}}(u)$ given a set of material-dependent parameters $t_0$, $\mathrm{K}$, $\Delta$, $\beta$ and a strain $\epsilon$. Lattice dynamics being neglected, we write the lattice contribution, which accounts for the displacement of the atoms with respect to their position in a uniformly spaced chain:
\begin{equation}
    \label{eq:def_EL}
    E_{\mathrm{L}}(u) = \frac{1}{2} \mathrm{K} \left( \delta r_A - \delta r_B \right)^2 + \frac{1}{2} \mathrm{K} \left( \delta r_B - \delta r_A \right)^2 = \frac{1}{4} \mathrm{K} a^2(\epsilon) u^2(\epsilon),
\end{equation}
where we used {Equation (\ref{eq:def_u})} and $\mathrm{K}$ is an elastic constant term.

\begin{figure}[!htb]
    \centering
    \includegraphics[width=0.5\textwidth]{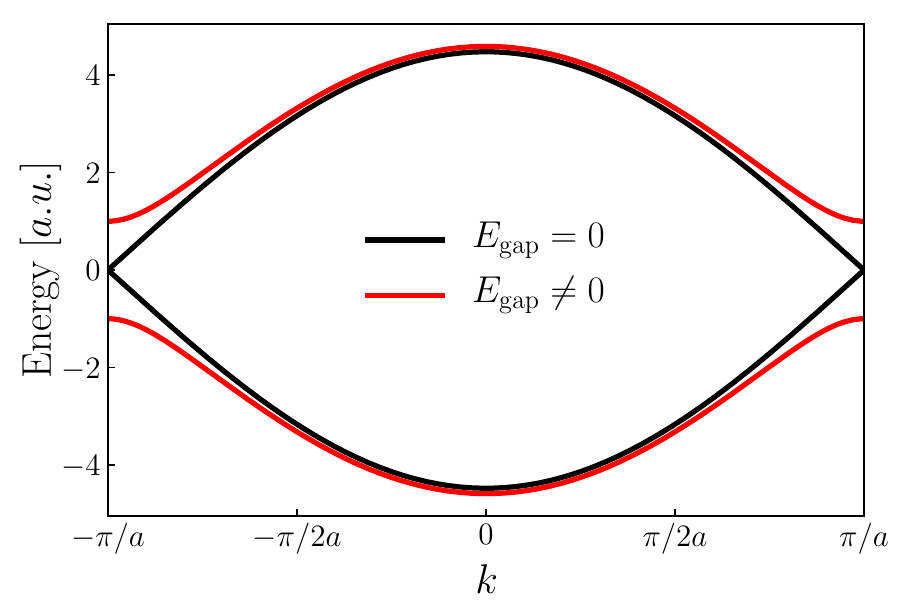}
    \caption{Electronic bands structure of the model. If $E_{\mathrm{gap}}=0$, the bands have linear dispersion at the edges of the Brillouin zone, reminding the behaviour of graphene. To describe the properties of conjugated polymers, we assume that only the lower energy bands are occupied with electrons.}
    \label{fig:bands}
\end{figure}

To compute the electronic energy per unit cell $E_{\mathrm{e}}(u)$, we imagine the linear chain as made of $N$ copies of the unit cell and we adopt periodic boundary conditions. This allows us to define a basis $\{\ket{\alpha, k}\}$ in the reciprocal $k$-space:
\begin{equation}
    \label{eq:Bloch_basis}
    \ket{\alpha, k} \equiv \sum_R \frac{\e^{\iu \left(r_{\alpha} + R \right) k}}{\sqrt{N}}\ket{\alpha R},
\end{equation}
where $k$ is defined over the first Brillouin zone, namely 
\begin{equation}
    k = \frac{n}{N} \frac{2\pi}{a}, \;\;\; n=0,\pm\frac{1}{2},\pm 1,\dots,\pm \frac{N}{2},
\end{equation}
and it holds
\begin{equation}
    \Braket{\alpha', k' | \alpha k} = \delta_{\alpha,\alpha'}\delta_{k,k'} .
\end{equation}
From {Equation (\ref{eq:def_Delta}), (\ref{eq:def_t1_t2})} and {(\ref{eq:Bloch_basis})} we obtain the matrix elements of the electronic Hamiltonian in the reciprocal space:
\begin{align}
    \Braket{A, k|H_{\mathrm{e}}|A, k'} &= -\Delta \delta_{k,k'}, \\
    \Braket{B, k|H_{\mathrm{e}}|B, k'} &= \Delta \delta_{k,k'}, \\
    \Braket{A, k|H_{\mathrm{e}}|B, k'}
    \label{eq:def_Tk}
    &=-t_1\e^{\iu\frac{a}{2}k} \e^{\iu k\left(\delta r_B - \delta r_A\right)} \delta_{k,k'} - t_2\e^{-\iu\frac{a}{2}k} \e^{\iu k\left(\delta r_B - \delta r_A\right)} \delta_{k,k'} \equiv -T \delta_{k,k'}, \\
    \Braket{B, k|H_{\mathrm{e}}|A, k'} &= \Braket{A, k'|H_{\mathrm{e}}|B, k}^*=-T^*\delta_{k,k'} .
\end{align}
The above results allow to write in the reciprocal space basis a $2\times2$ electronic Hamiltonian matrix $H_{\mathrm{e},k}$ for each $k$-point:
\begin{equation}
    \label{eq:def_Hk}
    H_{\mathrm{e},k} =
    \begin{pmatrix}
    -\Delta & -T \\
    -T^* & \Delta
    \end{pmatrix}.
\end{equation}
Diagonalising the matrix of {Equation (\ref{eq:def_Hk})} we find the two eigenvalues
\begin{equation}
    \label{eq:def_eigenvalues}
    \varepsilon_k^{\pm} = \pm \sqrt{\Delta^2 + |T|^2}
\end{equation}
with the respective eigenvectors, namely the Bloch wave-functions
\begin{equation}
    \label{eq:def_eigenvectors}
    \ket{\psi_k^{\pm}} = 
    \begin{pmatrix}
        \frac{\pm\left(\varepsilon_k^{\pm} - \Delta\right)}{\sqrt{\left( \varepsilon_k^{\pm} - \Delta \right)^2 + |T|^2}} \\
        \frac{-T^*}{\sqrt{\left( \varepsilon_k^{\pm} - \Delta \right)^2 + |T|^2}}
    \end{pmatrix},
\end{equation}
and using {Equation (\ref{eq:def_t_epsilon_bis}), (\ref{eq:def_deltat_epsilon})} and {\ref{eq:def_Tk}} we have that
\begin{equation}
    \label{eq:def_T2}
    |T|^2 = 4t^2\cos^2k\frac{a}{2}+4\delta t^2\sin^2k\frac{a}{2}.
\end{equation}
For each $k$-point, {Equation (\ref{eq:def_eigenvalues})} allows to distinguish between a lower and a higher energy level. As shown in {Figure \ref{fig:bands}}, the chain present two energy bands. To describe the $\pi-$orbitals of conjugated orbitals, we assume that only the lower energy band is filled with electrons and will hereafter refer to it as the occupied band, in contrast with the unoccupied higher energy band. From {Equation (\ref{eq:def_eigenvalues})} we obtain the value of the energy gap $E_{\mathrm{gap}}$ between the occupied and unoccupied band: 
\begin{equation}
    \label{eq:def_Egap}
    E_{\mathrm{gap}}=\sqrt{(4\delta t)^2 + (2\Delta)^2}.
\end{equation}

\begin{figure}[!htb]
    \centering
    \includegraphics[width=0.7\textwidth]{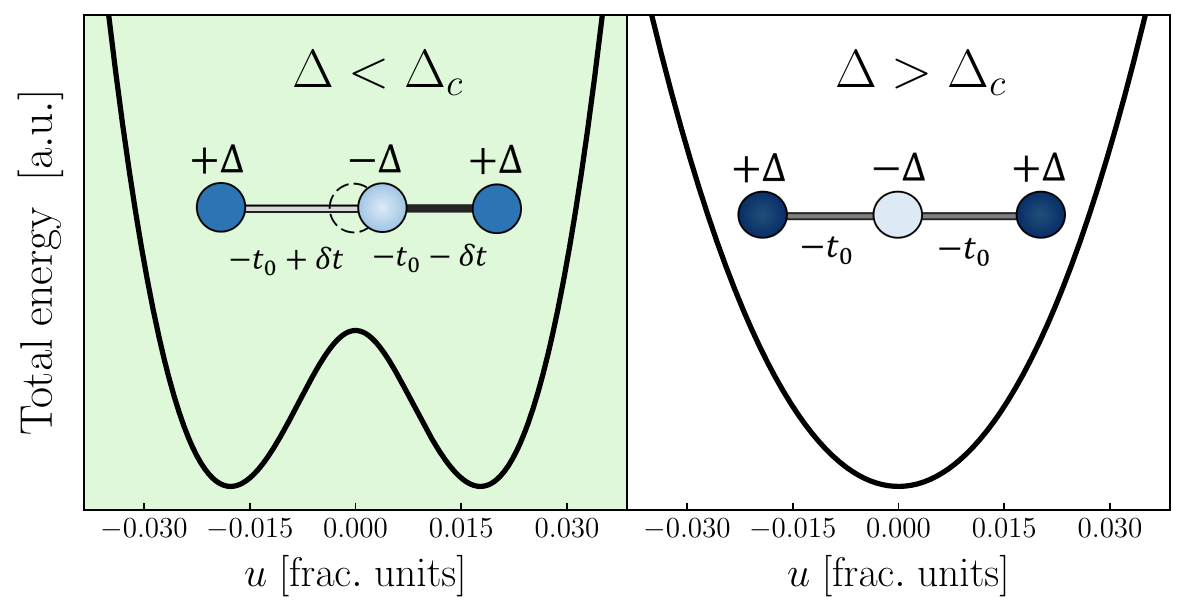}
    \caption{Behaviour of the total energy per unit cell $E_{\mathrm{tot}}$ with respect to the fractional coordinate $u$. If $\Delta\leq\Delta_{\mathrm{c}}$, there are two symmetric minima $\overline{u}\neq0$ and the equilibrium structure present a bond length alternation. If $\Delta>\Delta_{\mathrm{c}}$, there is only one minimum in $\overline{u}=0$ and the atoms are equidistant.}
    \label{fig:Etot_vs_u}
\end{figure}

We consider the contribution of all the occupied states to the electronic energy per unit cell and in particular it holds
\begin{equation}
    \label{eq:def_Ee}
    E_{\mathrm{e}} = \frac{2}{N}\sum_k \varepsilon_k^{-} = -\frac{2}{N} \sum_k \sqrt{\Delta^2 + 4t^2\cos^2\frac{ka}{2} + 4\delta t^2\sin^2\frac{ka}{2}}
\end{equation}
where the factor of $2$ accounts for the spin degeneracy. In the limit of an infinite linear chain, namely for $N\to+\infty$, the eigenvalues and the eigenvectors' coefficients become continuous functions of $k$ and the sum becomes an integral over the first Brillouin zone:
\begin{equation}
    \label{eq:def_Ee_bis}
    E_{\mathrm{e}} = -2 a\int_{-\pi/a}^{\pi/a} \frac{\dd k}{2\pi} \sqrt{\Delta^2 + 4t^2\cos^2\frac{ka}{2} + 4\delta t^2\sin^2\frac{ka}{2}} .
\end{equation}
Finally, using Equation (\ref{eq:def_deltat_epsilon_bis}), (\ref{eq:def_EL}) and (\ref{eq:def_Ee_bis}), we write the total energy per unit cell as a function of the fractional coordinate $u$:  
\begin{equation}
    \label{eq:def_Etot}
    E_{\mathrm{tot}}(u) = \frac{1}{4} \mathrm{K} u^2 a(\epsilon)^2 - 2 a\int_{-\pi/a}^{\pi/a} \frac{\dd k}{2\pi}\sqrt{\Delta^2+4t^2(\epsilon)\cos^2\frac{ka}{2} + 4\beta^2t_0^2(1+\epsilon)^2u^2\sin^2\frac{ka}{2} } .
\end{equation}
We can now study the structural properties of the system, encompassed in the optimal displacement $\overline{u}$, defined as the one which minimises $E_{\mathrm{tot}}(u) = E_{\mathrm{L}}(u) + E_{\mathrm{e}}(u)$ given a set of material-dependent parameters $t_0$, $\mathrm{K}$, $\Delta$, $\beta$ and a strain $\epsilon$. With the substitution $z=ka/2$ and exploiting the parity of the integrand in {Equation (\ref{eq:def_Etot})}, the first derivative of the total energy with respect to $u$ reads:
\begin{equation} \label{eq:def_dEdu}
    \frac{\dd E_{\mathrm{tot}}}{\dd u} = \frac{1}{2}\mathrm{K}a^2(\epsilon) u - \frac{1}{\pi}\int_{-\pi/2}^{0}\dd z \frac{ 16\beta^2t_0^2(1+\epsilon)^2 u \sin^2 z }{ \sqrt{\Delta^2+4t^2(\epsilon)\cos^2 z + 4\beta^2t_0^2(1+\epsilon)^2u^2\sin^2 z } }.
\end{equation}
One of the stationary point of {Equation (\ref{eq:def_dEdu})} is in $u=0$, whereas the others are the solutions of the following equation in $u$:
\begin{equation}
    \mathrm{K}a^2(\epsilon)\pi = \int_{-\pi/2}^{0} \dd z \frac{ 32\beta^2t_0^2(1+\epsilon)^2 \sin^2 z }{ \sqrt{\Delta^2+4t^2(\epsilon)\cos^2 z + 4\beta^2t_0^2(1+\epsilon)^2u^2\sin^2 z } } .
\end{equation}
To solve this integral, we expand the functions $\sin z$ and $\cos z$ at the lower edge of the Brillouin zone, namely in $k=-\pi/a$, in a similar fashion to the conic approximation in graphene. By doing so, we arrive at 
\begin{equation}
    \label{eq:def_u2}
    \Delta^2 + 4\beta^2t_0^2(1+\epsilon)^2u^2 = \frac{t_0^2(1-\beta\epsilon)^2\pi^2}{\sinh^2\frac{\mathrm{K}a_0^2\pi(1-\beta\epsilon)}{8\beta^2t_0}}. 
\end{equation}
Supposing that $\Delta$ is the parameter which guides the transition, we define
\begin{equation}
    \label{eq:def_DeltaC}
    \Delta_{\mathrm{c}} = \frac{t_0(1-\beta\epsilon)\pi}{\sinh\left(\frac{\mathrm{Ka_0^2\pi(1-\beta\epsilon)}}{16\beta^2t_0}\right) }.
\end{equation}
As we are interested in the real solutions only, the above equations tells that if $\Delta \leq \Delta_{\mathrm{c}}$, the solution of {Equation (\ref{eq:def_u2})} provides two symmetric stationary points for $E_{\mathrm{tot}}(u)$. Indicating with $\overline{u}$ the value which minimises $E_{\mathrm{tot}}(u)$, it is straightforward to verify, e.g. computing the second order derivative of $E_{\mathrm{tot}}(u)$, that 
\begin{equation}
    \begin{cases}
        \overline{u} \propto |\Delta - \Delta_{\mathrm{c}}|^{1/2}, &\; \mathrm{if} \; \Delta \leq \Delta_{\mathrm{c}} \\
        \overline{u} = 0, &\; \mathrm{if} \; \Delta > \Delta_{\mathrm{c}}.
    \end{cases}
\end{equation}
Moreover, it is also immediate to verify that the second order derivative of $E_{\mathrm{tot}}(u)$ computed at $u(\Delta_{\mathrm{c}})=0$ is a saddle point.
These results show that the chain undergoes a second order phase transition in $\Delta$ with $\overline{u}$ as order parameter: when $\Delta>\Delta_{\mathrm{c}}$, the atoms are equidistant ($\overline{u}=0$), while for $\Delta\leq\Delta_{\mathrm{c}}$ the chain displays a bond length alternation that breaks the inversion symmetry of the cell. {Figure \ref{fig:Etot_vs_u}} displays two representative energy profiles, one for each phase, obtained from {Equation (\ref{eq:def_Etot})}: the two minima of a double-well energy landscape in the distorted phase collapse into a single minimum when $\Delta\geq\Delta_{\mathrm{c}}$, i.e., when the local maximum at $u=0$ turns into a global minimum, signature of the second order transition.
It is interesting to notice that the left-hand side of {Equation (\ref{eq:def_u2})} is equal to $E^2_{\mathrm{gap}}/4$, implying that in the distorted phase, as $\Delta$ varies, $\overline{u}(\Delta)$ varies in a way that keeps the energy gap constant. Stated in other words, the knowledge of the energy gap gives also information on the phase diagram of the system and vice versa. However, 
the parameter $\Delta$ is not the only handle available to drive the transition, that may be tuned by other model parameters at fixed $\Delta$. For instance, it is reasonable that a sufficiently large e-ph coupling may induced bond dimerization in the gapped chain at finite $\Delta$. The optimal $\bar{u}\neq 0$ as a function of $\beta$ is given in closed form in {Equation (\ref{eq:def_u2})}. As the e-ph coupling constant enters in the hyperbolic function, an explicit expression for $\bar{u}$ can be derived by assuming $\mathrm{K}a_0^2\pi(1-\beta\epsilon) \ll 8\beta^2t_0$, which allows to retain only the lowest order term of the Taylor expansion of $\sinh(x)$. With this hypothesis, and defining the term 
\begin{equation}
    \beta_{\mathrm{c}}^2 = \frac{\mathrm{K}a_0^2}{8t_0^2}\Delta,
\end{equation}
in analogy with the previous case, we obtain 
\begin{equation}
    \begin{cases}
        \overline{u} \propto |\beta - \beta_{\mathrm{c}}|^{1/2}, &\; \mathrm{if} \; \beta > \beta_{\mathrm{c}} \\
        \overline{u} = 0, &\; \mathrm{if} \; \beta \leq \beta_{\mathrm{c}} .
    \end{cases}
\end{equation}
This result shows that another way to control the structure of the chain in the Rice-Mele model, namely given a finite $\Delta\neq0$, is through the electron-phonon coupling term $\beta$. {  Both the behaviours of otpimal $\overline{u}_0(\Delta)$ and $\overline{u}_0(\beta)$ evaluated at zero strain are shown in Figure \ref{fig:polarization}.}

\section*{Electronic polarization and Thouless pump}
\begin{figure}[!htb]
    \begin{minipage}[c]{1.\linewidth}
    \centering
    \includegraphics[width=0.32\textwidth]{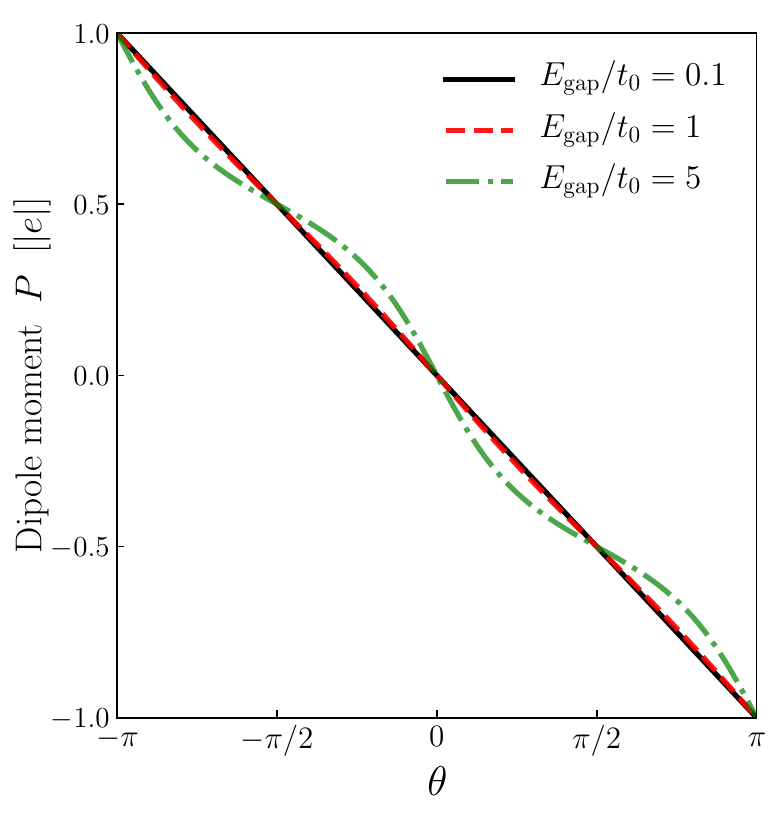}
    \includegraphics[width=0.63\textwidth]{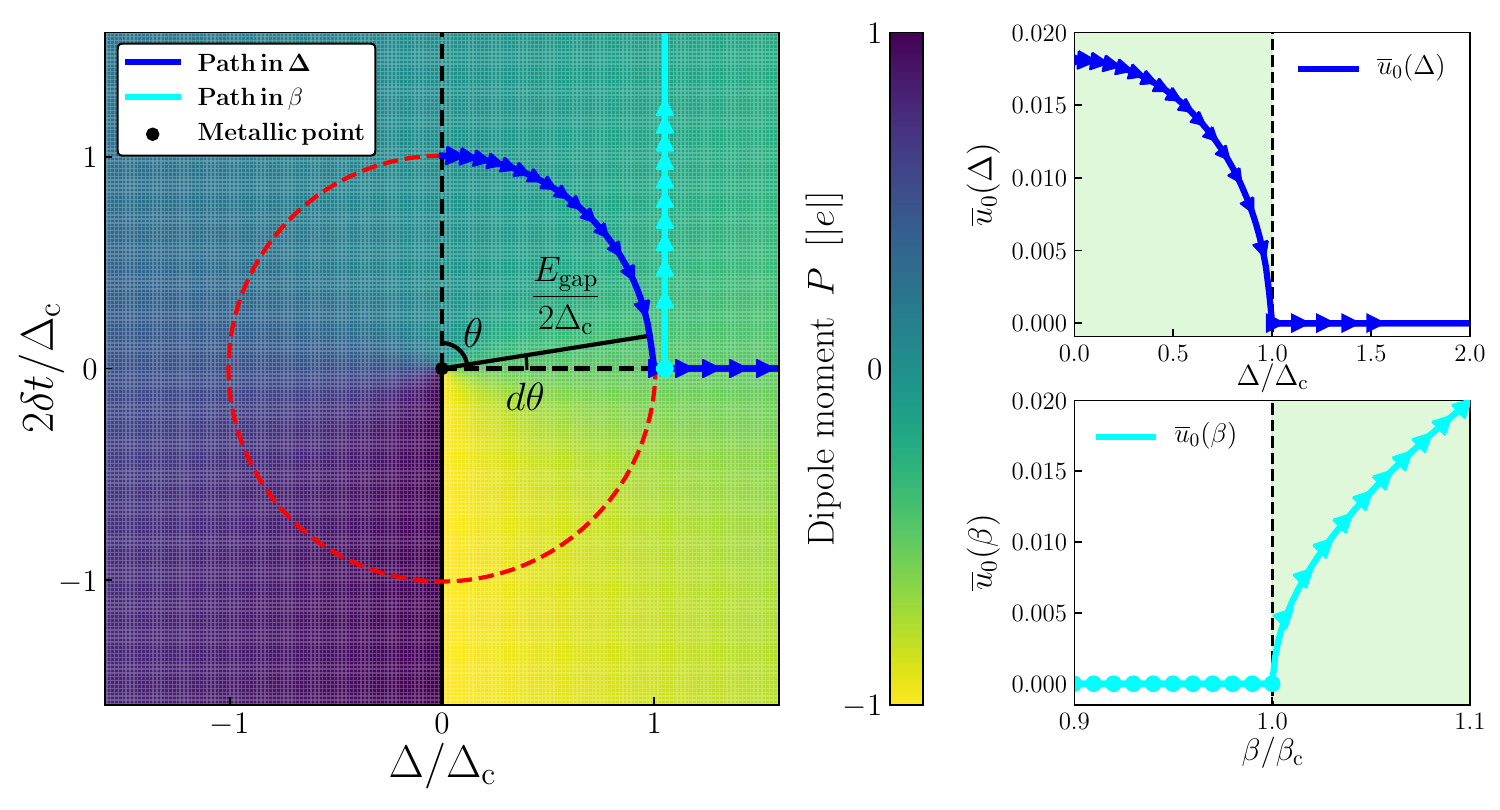}
    \caption{On the left panel, behaviour of the dipole moment per unit cell $P$ with respect to $\theta$ for different values of $E_{\mathrm{gap}}/t_0$. As expected, in the small-gap limit ($E_{\mathrm{gap}} \lesssim t_0$) we observe the linear behaviour $P\propto-\theta$ predicted in Equation (\ref{eq:def_P_theta}). On the central panel, the evolution of $P$ is displayed in the 2D parametric $(\Delta, \delta t)-$space, where each point correspond to a physical realization of the model. An adiabatic evolution along a loop containing the origin of the axes -- which correspond to a metallic system -- pumps out a quantised charge of $-2|e|$. In the 2D space on the central panel it is possible to appreciate the geometric argument that yields Equation (\ref{eq:Zeff_vs_gap}). Being the polarization proportional to the angle $\theta$ spanning the parametric ($\Delta,\delta t$)-space along a circumference with radius $E_{\mathrm{gap}}$, as $\delta t\propto u$, it follows that in the undimerized phase $E_{\mathrm{gap}} d\theta\propto du$, hence $Z^*=\partial P/\partial u \propto\partial \theta/\partial u\propto 1/E_{\mathrm{gap}}$. Finally, it is interesting to visualize the phase transitions of $\overline{u}_0(\Delta)$ and $\overline{u}_0(\beta)$, displayed separately in the two right panels, also as paths in the 2D parametric space. Being $\Delta$ and $\beta$ the independent parameters and $\delta t\propto\beta\overline{u}$, one finds that the transition in $\Delta$ from the dimerized ($\overline{u}_0\neq0$, hence $\delta t\neq0$) to the undimerized ($\overline{u}_0=0$, hence $\delta t=0$) phase, happens on a circumference with fixed $E_{\mathrm{gap}}$, whereas the evolution with respect to $\beta$ in the dimerized phase corresponds to a vertical line, the undimerized phase corresponding to a single point at fixed $\Delta\neq 0$ and $\delta t=0$. Arrows on the two paths in the 2D space correspond, respectively, to the arrows in the right panels whereas all the dots of the lower right panel correspond to a single point in the 2D space in $(\Delta=1.05\Delta_{\mathrm{c}},\delta t=0)$.}
    \label{fig:polarization}
    \end{minipage}
\end{figure}

In this section, we derive the expression of the dipole moment per unit cell $P$ as obtained within the modern theory of polarization\cite{king1993theory}. In this framework, the wave function of the occupied states is required to be continuous at the edges of the Brillouin zone. To satisfy the requirement, we multiply  the wave function for the occupied states of {Equation (\ref{eq:def_eigenvectors})} by a phase factor and define 
\begin{equation}
    \label{eq:def_psik_occ}
    \ket{\psi_k^{\mathrm{occ}}} \equiv \e^{-\iu r_A k} \ket{\psi_k^{-}} 
\end{equation}
with $r_A = - a/4 + \delta r_{A}$. The dipole moment per unit cell $P$ is defined in term of the Berry phase\cite{berry1984quantal} $\varphi$ as 
\begin{equation}
    \label{eq:def_polarization}
    P=\frac{-2|e|}{2\pi}\varphi
\end{equation}
where $e$ is the electron's charge, the factor $2$ at the numerator accounts for the spin degeneracy and the Berry phase $\varphi$ is defined as
\begin{equation}
    \label{eq:def_berry_phase}
    \varphi = \iu\int_{-\pi /a}^{\pi /a} \dd k \Braket{\widetilde{\psi}_k^{\mathrm{occ}} | \frac{\dd \widetilde{\psi}_k^{\mathrm{occ}}}{\dd k} }.
\end{equation}
We indicated with $\ket{\widetilde{\psi}_k^{\mathrm{occ}}}$ the periodic part of the wave function $\ket{\psi_k^{\mathrm{occ}}}$ and from {Equation (\ref{eq:Bloch_basis}), (\ref{eq:def_eigenvectors})} and {(\ref{eq:def_psik_occ})} it holds
\begin{equation}
    \label{eq:def_psik_occ_periodic}
    \ket{\widetilde{\psi}_k^{\mathrm{occ}}} = \e^{-\iu r_A k} \sum_R\sum_{\alpha}c_{\alpha, k}\ket{\alpha R},
\end{equation}
where 
\begin{align}
    \label{eq:coeff_Ok}
    c_{A,k} &= \frac{\Delta - \varepsilon_k^{-}}{\sqrt{\left( \Delta - \varepsilon_k^{-} \right)^2 + |T|^2}}, \\
    c_{B,k} &= \frac{-T^*}{\sqrt{\left(\Delta - \varepsilon_k^{-} \right)^2 + |T|^2}}.
\end{align}
As the scalar product in {Equation (\ref{eq:def_berry_phase})} is taken over a single unit cell, without loss of generality we consider only the contribution for $R=0$ in Equation (\ref{eq:def_psik_occ_periodic}) and obtain
\begin{equation}
    \label{eq:computed_berrry_phase}
    \varphi = \frac{2\pi}{a}r_{A} +\int_{\pi/a}^{\pi/a} \dd k \frac{\left(\delta r_B - \delta r_A\right)|T|^2 + 2t\delta t a}{(\Delta - \varepsilon_k^-)^2 + |T|^2}.
\end{equation}
From the gap {Equation (\ref{eq:def_Egap})}, we notice that the insulating/metallic character of the system can be visualised in a 2D parametric $(\Delta, \delta t)-$space, where the origin of the axes correspond to a metallic system with $E_{\mathrm{gap}}=0$, while every other point correspond to an insulating system with $E_{\mathrm{gap}}\neq0$. In this space, we define a parameter $\theta$ that allows to identify each point with the polar coordinates $(E_{\mathrm{gap}},\theta)$, with the change of coordinates defined by
\begin{align}
    \label{eq:def_sintheta}
    2\Delta &= E_{\mathrm{gap}} \sin\theta \\
    \label{eq:def_costheta}
    4\delta t &= E_{\mathrm{gap}} \cos\theta.
\end{align}
Applying this change of coordinates in {Equation (\ref{eq:computed_berrry_phase})}, in the limit $E_{\mathrm{gap}} \ll t_0$ it can be demonstrated that it holds
\begin{equation}
    \label{eq:def_P_theta}
    P = -\frac{|e|}{\pi}\theta.
\end{equation}
This property can be appreciated in {Figure \ref{fig:polarization}}, where we compare, for different values of $E_{\mathrm{gap}}/t_0$, the behaviour with respect to $\theta$ of the dipole moment $P$, obtained computing {Equation (\ref{eq:def_polarization})} numerically without further approximations. 
As $\theta$ varies, so do the terms $\Delta$ and $\delta t$. In particular, a full rotation of $2\pi$ implies that the system has returned in its initial state, while the dipole moment $P$ has acquired a quantum of $-2|e|$. This properties is more general: if the system undergoes an adiabatic evolution along any loop enclosing the origin of the 2D space, a quantised charge is always pumped out. This phenomenon is known as adiabatic charge transport or Thouless’ pump\cite{thouless1983quantization} and the key ingredient is the presence of the metallic point in the domain enclosed by the loop: in this sense, it is an example of topological phenomenon. This peculiar topology-related property has consequences also in the behaviour of the effective charges of the system. Defined as the variation of polarization due to an atomic displacement, the effective charges are a measure of how rigidly the electronic charge distribution follows the displacement of the nuclei. In the model it holds 
\begin{equation}
    Z^{*} = \frac{\partial P}{\partial u}.
\end{equation}
With the definitions given above one can compute analytically the expression for $Z^*$, however it is possible to appreciate the main feature of its behaviour thanks to a geometric argument in the ($\Delta,\delta t$)-space. The effect of a small displacement on a system with a finite $\Delta\neq0$ is to span an angle $\dd\theta$ in the space, starting from $\theta=\pi/2$. Considering that $\delta t \propto u$ and that, with a suitable choice of the axis, the points on a circumference in this space correspond to systems with the same $E_{\mathrm{gap}}$, we have that $\dd\theta \propto u /E_{\mathrm{gap}}$. From {Equation (\ref{eq:def_P_theta})}, it holds $\dd P\propto\dd\theta$, hence we obtain
\begin{equation} \label{eq:Zeff_vs_gap}
    Z^* \propto \frac{1}{E_{\mathrm{gap}}},
\end{equation}
implying that as we get closer to origin of the axis ($E_{\mathrm{gap}}$ goes to $0$), even an infinitesimal atomic displacement causes a huge redistribution of the charge density. { This geometric argument can be appreciated in Figure \ref{fig:polarization}.}

\section*{Dynamical charges of a heteropolar diatomic molecule}
\begin{figure}[!htb]
    \begin{minipage}[c]{1.\linewidth}
    \centering
    \includegraphics[width=.7\textwidth]{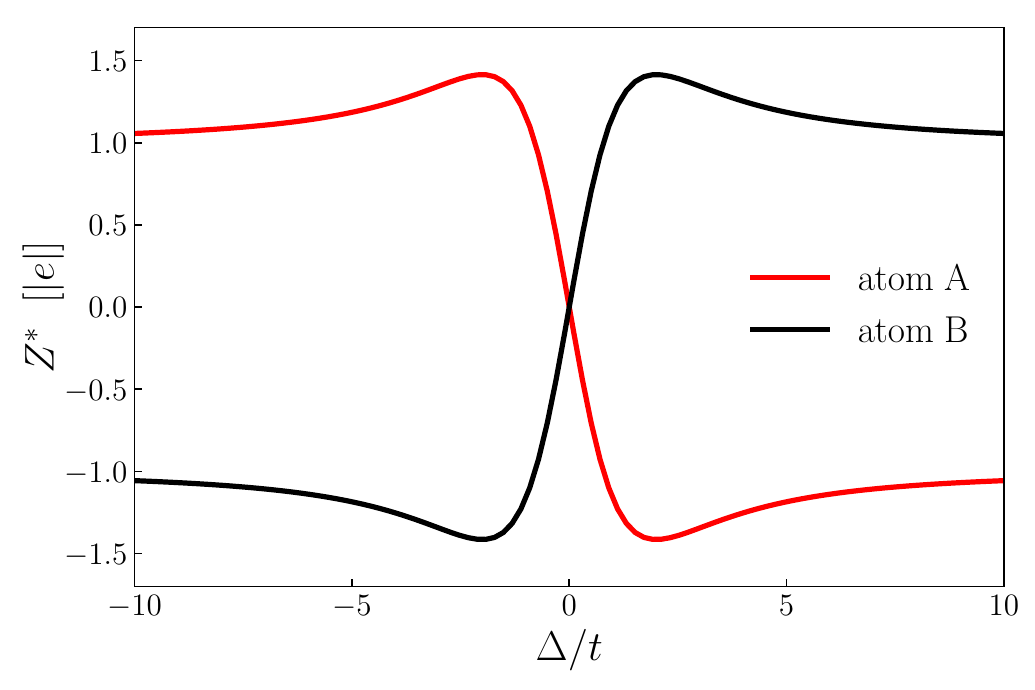}
    \caption{Dynamical charges of atoms $A$ and $B$. Tuning the ionic/covalent character of the bond through the ratio $\Delta/t$ allows for a maximum enhancement of $\sim 1.5$ times the valence charge of $1|e|$, in stark contrast with the enhancement of up to $30$ times we find in the chain.}
    \label{fig:Zdyn}
    \end{minipage}
\end{figure}

In this Section we compute the dynamical charge of a heteropolar diatomic molecule along the lines discussed in the appendix of Ref. \cite{Ghosez.PhysRevB.58.6224}, in order to contrast our predicted topological enhancement of effective charge and the contribution of the mixed ionic-covalent character to ‘‘anomalous'' Born effective charges. Let's consider a diatomic molecule with two monovalent atoms $A$ and $B$ positioned along the $x$-axis at a distance $d_{AB}=R_B-R_A>0$. In a LCAO tight-binding approach, the electronic Hamiltonian $H_{\mathrm{e}}$ reads
\begin{equation}\label{eq:def_He_dimero}
    H_{\mathrm{e}} = E_{A}\ket{A}\bra{A} + E_{B}\ket{B}\bra{B} - t(d_{AB}) \left( \ket{A}\bra{B} + \ket{B}\bra{A} \right)
\end{equation}
where $t(d_{AB})$ is the hopping energy between the atoms, that will in general depend on some power of the inverse distance. For the sake of clarity and without loss of generality, we assume $t(d_{AB}) \propto 1/d_{AB}^2.$ We can find the occupied electronic orbital $\ket{\psi^{\mathrm{occ}}_{\mathrm{e}}}$ diagonalizing $H_{\mathrm{e}}$:
\begin{align}
\label{eq:def_psi_occ}
    \ket{\psi^{\mathrm{occ}}_{\mathrm{e}}} 
    &= \sqrt{\frac{1 + x}{2}} \ket{A} + \sqrt{\frac{1 - x}{2}} \ket{B} 
\end{align}
with $x=\Delta/\sqrt{\Delta^2 + 4t^2}$ and $\Delta = E_B - E_A$. This allows to define the dipole moment $D(d_{AB})$ of the molecule as 
\begin{align} 
\label{eq:def_dipole}
    D(d_{AB}) 
    &= R_A Z_A^v + R_B Z_B^v - 2 \bra{\psi^{\mathrm{occ}}_{\mathrm{e}}} \hat{r} \ket{\psi^{\mathrm{occ}}_{\mathrm{e}}}
\end{align}
where $Z_A^v=Z_B^v=+1|e|$ are the valence charges of atom A and B, respectively, and $\hat{r}$ is the position operator, whose matrix elements are well defined in an isolated molecule. The dynamical charge $Z^*_{\alpha}$ of atom $\alpha=A,B$ is defined as the derivative of $D(d_{AB})$ with respect to the atomic displacement $R_{\alpha}$, and using all the above definitions we obtain (in units of $\vert e\vert$)
\begin{align}
    Z^*_{\alpha} 
    &= \frac{\partial D(d_{AB})}{\partial R_{\alpha}} \label{eq:def_Zdyn}\\
    &= \frac{(-1)^{i_\alpha}\Delta}{\sqrt{\Delta^2 + 4t^2}}\left( 1 + \frac{8t^2}{\Delta^2 + 4t^2} \right) \label{eq:def_Zdyn1}\\
    &= \frac{2X}{\sqrt{1+4X^2}}\left(1+\frac{2}{1+4X^2} \right)\label{eq:def_Zdyn2}
\end{align}
where $i_{A}=1$ and $i_{B}=2$. The above equations tell us that acting on the ionic-covalent character of the bond, accounted for by the term $X=\Delta/t$, one can tune the values of the dynamical charges. Albeit the expression of Eq. \ref{eq:def_Zdyn1} apparently reminds the dependence on the gap of the effective charge in the undimerized 1D chain provided in Eq. \ref{eq:Zeff_vs_gap}, as it inversely depends on the difference between molecular energy levels, from Eq. \ref{eq:def_Zdyn2} it is clear that the effective charge is always limited and shows no diverging behaviour. The dynamical charges of atoms $A$ and $B$, which obeys the charge-neutrality sum rule $Z^*_A + Z^*_B = 0$, are displayed in Figure \ref{fig:Zdyn} as a function of the ionic-covalent character $X=\Delta/t$. The maximum enhancement is indeed found for finite values of $X$, i.e., arising from the mixed ionic-covalent character of the bond, reaching however a finite value that is roughly $\sim1.5$ the nominal value, consistently with the enhancement reported in \cite{Ghosez.PhysRevB.58.6224}. We contrast this result with the enhancement of up to $30$ times the nominal charge that we find in the 1D chain and highlight the differences with the analytical behaviour found in the model, namely with Equation (\ref{eq:Zeff_vs_gap}).

\clearpage
\section*{Numerical calculations}
\subsection*{Structural information}
\begin{figure}[!htb]
    \centering
    \begin{minipage}[c]{1.0\linewidth}
    \centering
    \includegraphics[width=.6\textwidth]{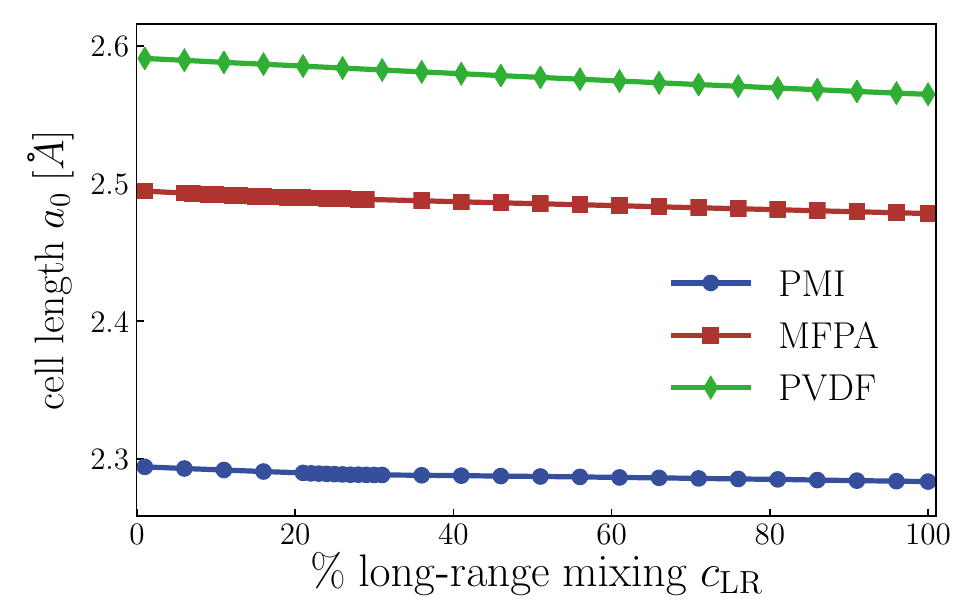}
    \end{minipage}
    \caption{Values of the cell length $a_0$ obtained through a cell relaxation for different values of the long-range mixing parameter $c_{\mathrm{LR}}$.}
    \label{fig:cell_length_relax}
\end{figure}

\begin{figure}[!htb]
    \centering
    \begin{minipage}[c]{1.0\linewidth}
    \centering
    \includegraphics[width=.6\textwidth]{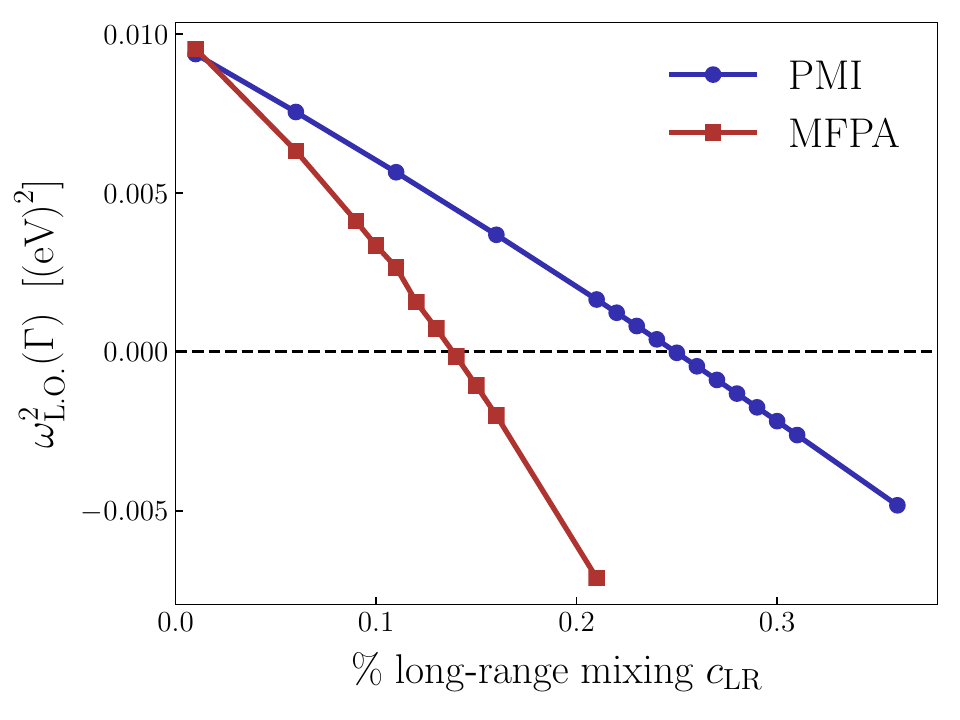}
    \end{minipage}
    \caption{Behaviour of the LO eigenvalue $\omega^2_{\mathrm{LO}}$, calculated at $\Gamma$ in the undistorted configuration, for different values of the long-range mixing parameter $c_{\mathrm{LR}}$. The change from positive to negative values, signalling a dynamical instability of the system toward the dimerized phase, corresponds to the change of curvature of the total energy at $u=0$ shown in Figure \ref{fig:Etot_vs_u}.}
    \label{fig:omega2_vs_cLR}
\end{figure}

\begin{figure}[!htb]
    \begin{minipage}[c]{1.0\linewidth}
    \centering
    \includegraphics[width=.45\textwidth]{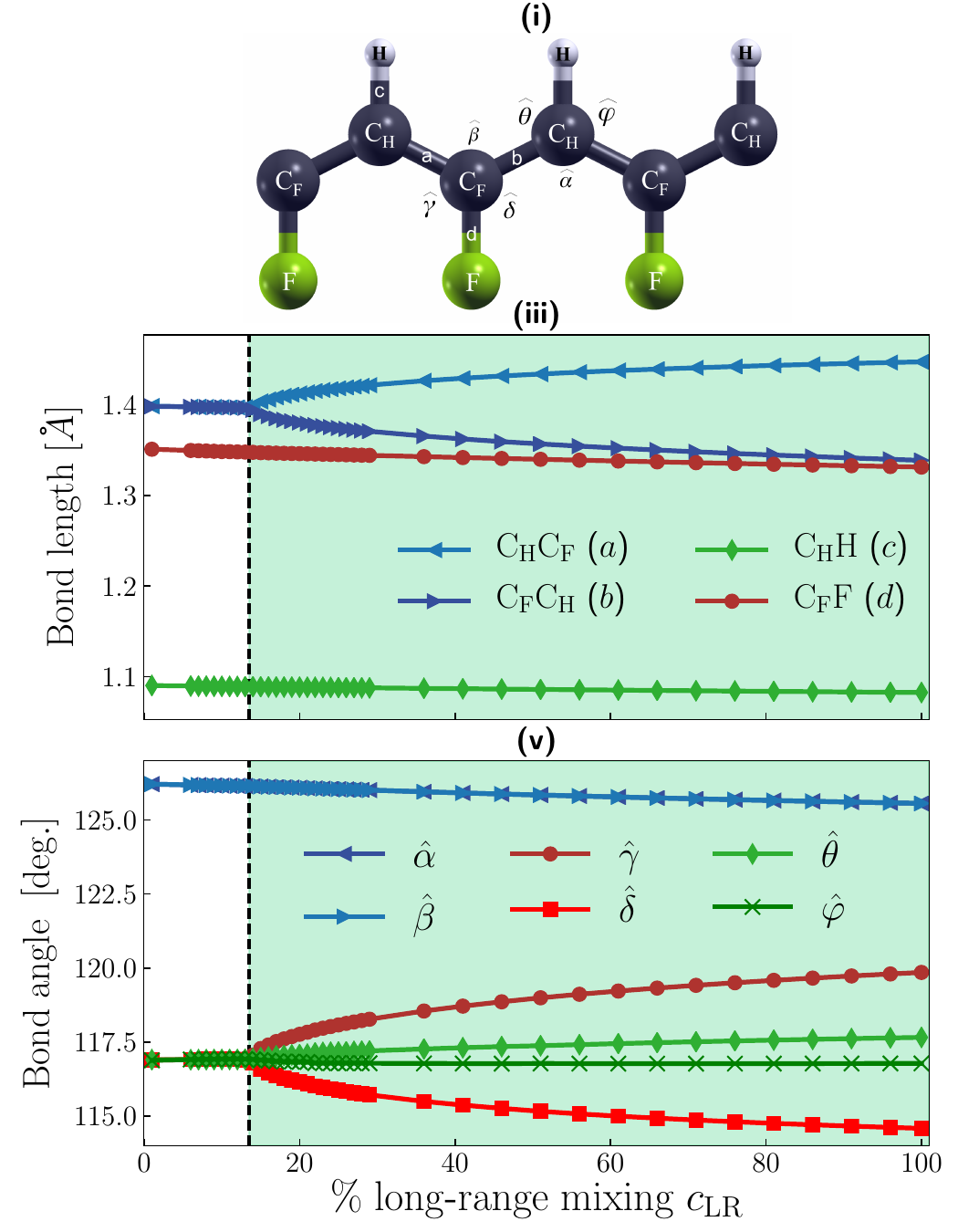}
    \includegraphics[width=.45\textwidth]{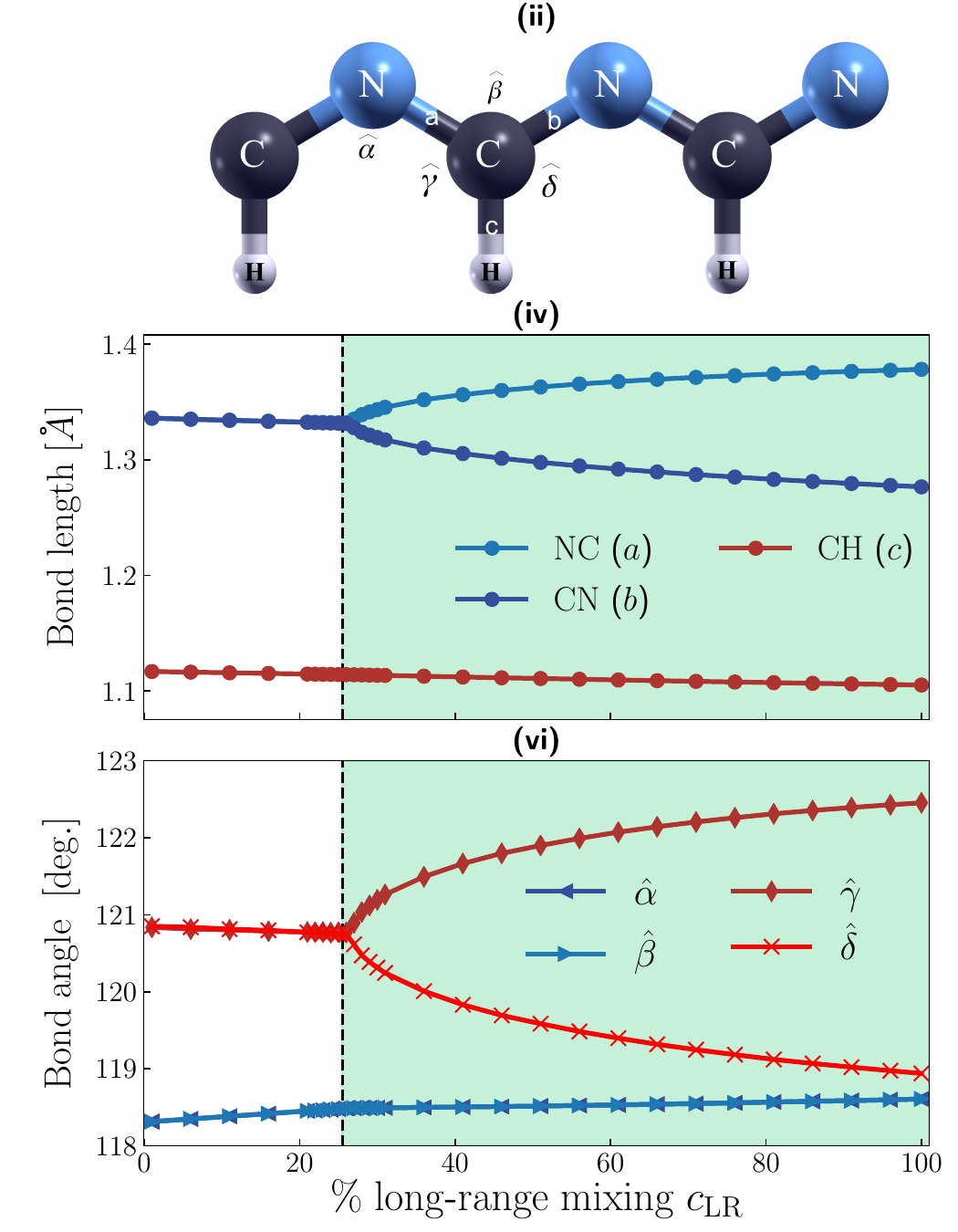}
    \caption{View of the undimerized structure of \textbf{(i)} mono-fluorinated polyacetylene (MFPA) and \textbf{(ii)} polymethineimine (PMI). In \textbf{(iii)} and \textbf{(iv)} for MFPA and PMI, respectively, we report the values of the bond length between the atoms of the unit cell, calculated with different long-range mixing parameter $c_{\mathrm{LR}}$, whereas in \textbf{(v)} and \textbf{(vi)} we report the values of the bond angles for MFPA and PMI, respectively. For MFPA, we use $\mathrm{C_H}$ and $\mathrm{C_F}$ to label carbon atoms bonded to hydrogen and fluorine, respectively. Bond angles are labeled as follows: $\hat\alpha$ and $\hat\beta$ denote the $\widehat{\mathrm{C_F C_H C_F}}$ ($\widehat{\mathrm{CNC}}$) and $\widehat{\mathrm{C_H C_F C_H}}$ ($\widehat{\mathrm{NCN}}$) bond angles, while $\hat\gamma$ and $\hat\delta$ label the $\widehat{\mathrm{C_H C_F F}}$ ($\widehat{\mathrm{NCH}}$) and $\widehat{\mathrm{FC_FC_H}}$ ($\widehat{\mathrm{HCN}}$) bond angles. For MFPA only, we also denote by $\hat\theta$ and $\hat\phi$ the $\widehat{\mathrm{C_FC_HH}}$ and $\widehat{\mathrm{HC_HC_F}}$ bond angles. As expected, the phase transition is signalled by the bond-length alternation as well as in the bifurcation of $\hat\gamma$ and $\hat\delta$ bond angles.}
    \label{fig:MFPA_PMI_structure}
    \end{minipage}
\end{figure}

\begin{figure}[!thb]
    \begin{minipage}[c]{.45\linewidth}
    \centering
    \includegraphics[width=1.\textwidth]{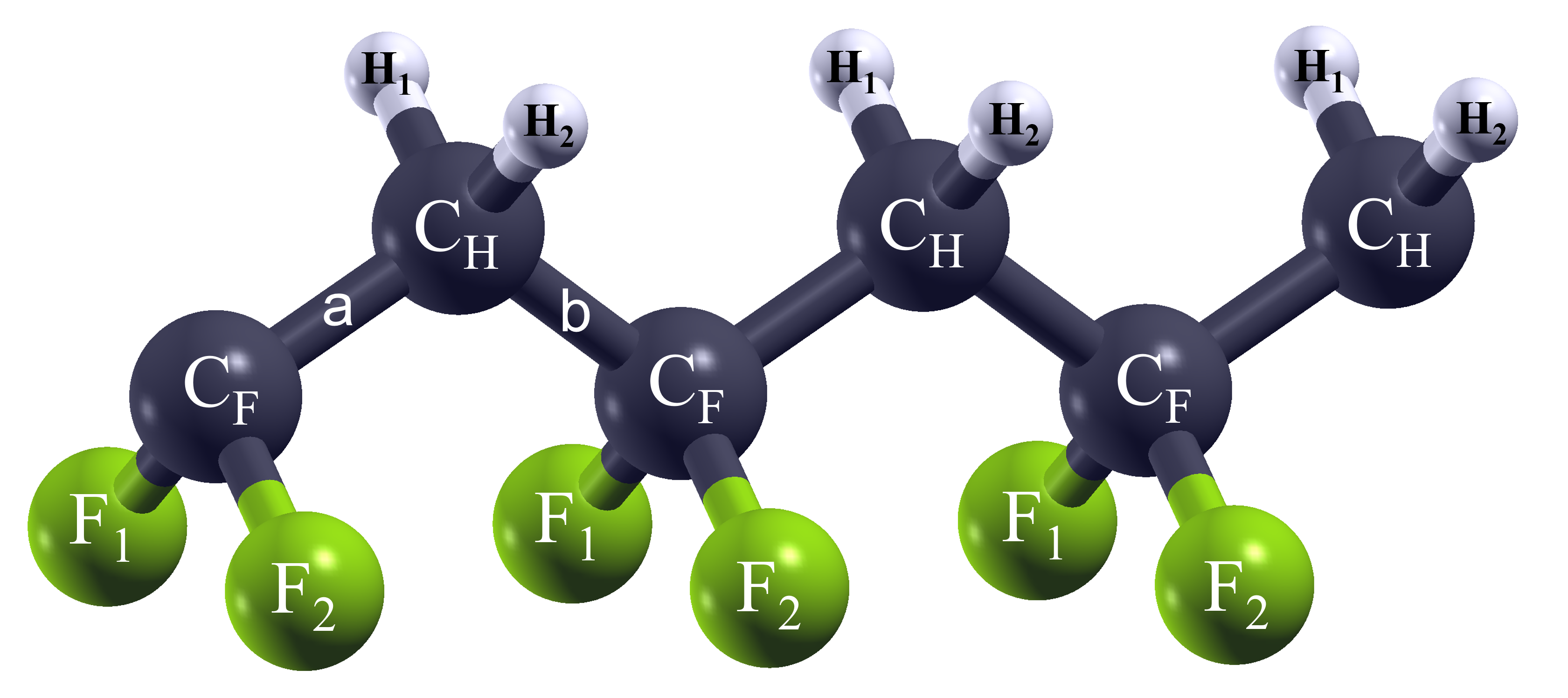}
    \end{minipage}
    \begin{minipage}[c]{.50\linewidth}
    \centering
    \includegraphics[width=1.\textwidth]{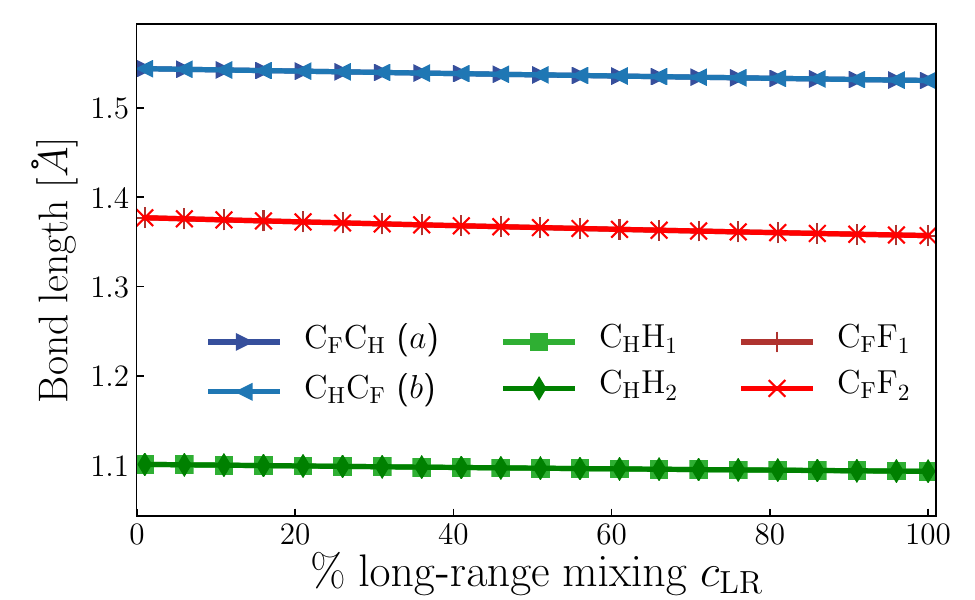}
    \end{minipage}
    \caption{On the left, view of the structure of polyvinylidene fluoride (PVDF). On the right, behaviour of the bond lengths for different values of $c_{\mathrm{LR}}$. As expected, the structure of PVDF depends very weakly on the long-range mixing parameter, this polymer being not conjugated.}
    \label{fig:PVDF_structure}
\end{figure}

For each value of the long-range mixing parameter $c_{\mathrm{LR}}$, a geometric optimisation was performed in order to obtain the equilibrium structures. The unit cell length $a_0$ has a very weak dependence on $c_{\mathrm{LR}}$ for all the polymers studied, as shown in {Figure \ref{fig:cell_length_relax}}. Signatures of the second order phase transition in MFPA and PMI are provided by { the behaviour of the eigenvalues of the dynamical matrix corresponding to the longitudinal optical (LO) mode driving the structural transition, calculated at $\Gamma$ in the undistorted configuration and shown in Figure \ref{fig:omega2_vs_cLR}, and} by the behaviour of the bond lengths and bond angles, reported in {Figure \ref{fig:MFPA_PMI_structure}}. Conversely, the structure of PVDF is not affected by changes in the $c_{\mathrm{LR}}$ parameter, as highlighted in {Figure \ref{fig:PVDF_structure}}.

\clearpage

\subsection*{Born effective charges of MFPA}
\vspace{-1cm}
\begin{table}[!htb]
    \centering
    \caption{For three representative values of the long-range mixing parameter $c_{\mathrm{LR}}$ (including the critical value where the chain dimerization occurs), the components of the effective charge tensors for each atom of the unit cell of MFPA are reported. The $Z^*_{xx}$ component of the carbon atoms reaches strikingly high values, hint of the anomalous polar response of the $\pi$-orbitals along the backbone chain. The sum rule $Z^*_{\mathrm{C_F}} + Z^*_{\mathrm{C_H}} + Z^*_{\mathrm{F}} + Z^*_{\mathrm{H}} = 0$ is also respected.\vspace{.4cm}}
    \label{tab:Zeff_MFPA}
    \begin{minipage}[c]{0.3\textwidth}
    \centering
    \begin{tabular}{cccc}
    \multicolumn{4}{c}{$\mathbf{c_\mathrm{\textbf{LR}} \boldsymbol{=} 1\boldsymbol{\%}}$} \\
    \hline \hline
    \multicolumn{4}{c}{$Z^*_{\mathrm{C_F}}$} \vspace{0.05cm} \\
    \hline
     & $x$ & $y$ & $z$ \\
     $x$ &  27.830 & -0.007 & 0.000 \\ 
     $y$ & -0.013  &  0.976 & 0.000 \\ 
     $z$ &  0.000  &  0.000 & 0.168 \\
    \vspace{-0.3cm} \\
    \hline \hline
    \multicolumn{4}{c}{$Z^*_{\mathrm{C_H}}$} \vspace{0.05cm} \\
    \hline 
     & $x$ & $y$ & $z$ \\
     $x$ & -25.533 & -0.001 &  0.000 \\ 
     $y$ &  0.013  & -0.104 &  0.000 \\ 
     $z$ &  0.000  &  0.000 & -0.197 \\ 
    \vspace{-0.3cm} \\
    \hline \hline
    \multicolumn{4}{c}{$Z^*_{\mathrm{F}}$} \vspace{0.05cm} \\
    \hline 
     & $x$ & $y$ & $z$ \\
     $x$ & -2.143 &  0.010 &  0.000 \\ 
     $y$ &  0.001 & -0.865 &  0.000 \\ 
     $z$ &  0.000 &  0.000 & -0.104 \\
    \vspace{-0.3cm} \\
    \hline \hline
    \multicolumn{4}{c}{$Z^*_{\mathrm{H}}$} \vspace{0.05cm} \\
    \hline 
     & $x$ & $y$ & $z$ \\
     $x$ & -0.154 & -0.002 & 0.000 \\ 
     $y$ &  0.000 & -0.007 & 0.000 \\ 
     $z$ &  0.000 &  0.000 & 0.132 \\
    \hline \hline
    \end{tabular}
    \end{minipage}
    %
    %
    \begin{minipage}[c]{0.3\textwidth}
    \centering
    \begin{tabular}{cccc}
    \multicolumn{4}{c}{$\boldsymbol{c_\mathrm{\textbf{LR}} = 13\% \, (\simeq \overline{c}_\mathrm{\textbf{LR}})}$} \\
    \hline \hline
    \multicolumn{4}{c}{$Z^*_{\mathrm{C_F}}$} \vspace{0.05cm} \\
    \hline 
     & $x$ & $y$ & $z$ \\
     $x$ &  30.497 & -0.015 & 0.000 \\ 
     $y$ & -0.029  &  0.991 & 0.000 \\ 
     $z$ &  0.000  &  0.000 & 0.185 \\
    \vspace{-0.3cm} \\
    \hline \hline
    \multicolumn{4}{c}{$Z^*_{\mathrm{C_H}}$} \vspace{0.05cm} \\
    \hline 
     & $x$ & $y$ & $z$ \\
     $x$ & -28.090 & -0.006 & 0.000 \\ 
     $y$ &  0.028  & -0.113 & 0.000 \\ 
     $z$ &  0.000  &  0.000 & -0.214 \\
    \vspace{-0.3cm} \\
    \hline \hline
    \multicolumn{4}{c}{$Z^*_{\mathrm{F}}$} \vspace{0.05cm} \\
    \hline 
     & $x$ & $y$ & $z$ \\
     $x$ & -2.240 &  0.028 & 0.000 \\ 
     $y$ &  0.001 & -0.876 & 0.000 \\ 
     $z$ &  0.000 &  0.000 & -0.106 \\
    \vspace{-0.3cm} \\
    \hline \hline
    \multicolumn{4}{c}{$Z^*_{\mathrm{H}}$} \vspace{0.05cm} \\
    \hline 
     & $x$ & $y$ & $z$ \\
     $x$ & -0.167 & -0.008 & 0.000 \\ 
     $y$ &  0.000 & -0.002 & 0.000 \\ 
     $z$ &  0.000 &  0.000 & 0.135 \\
    \hline \hline
    \end{tabular}
    \end{minipage}
    %
    %
    \begin{minipage}[c]{0.3\textwidth}
    \centering
    \begin{tabular}{cccc}
    \multicolumn{4}{c}{$\mathbf{c_\mathrm{\textbf{LR}} \boldsymbol{=} 100\boldsymbol{\%}}$} \\
    \hline \hline
    \multicolumn{4}{c}{$Z^*_{\mathrm{C_F}}$} \vspace{0.05cm} \\
    \hline 
     & $x$ & $y$ & $z$ \\
     $x$ &  3.969 & -0.901 & 0.000 \\ 
     $y$ & -0.334 &  1.100 & 0.000 \\ 
     $z$ &  0.000 &  0.000 & 0.217 \\
    \vspace{-0.3cm} \\
    \hline \hline
    \multicolumn{4}{c}{$Z^*_{\mathrm{C_H}}$} \vspace{0.05cm} \\
    \hline 
     & $x$ & $y$ & $z$ \\
     $x$ & -3.426 &  0.145 &  0.000 \\ 
     $y$ &  0.316 & -0.154 &  0.000 \\ 
     $z$ &  0.000 &  0.000 & -0.241 \\ 
    \vspace{-0.3cm} \\
    \hline \hline
    \multicolumn{4}{c}{$Z^*_{\mathrm{F}}$} \vspace{0.05cm} \\
    \hline 
     & $x$ & $y$ & $z$ \\
     $x$ & -0.576 &  0.088 &  0.000 \\ 
     $y$ &  0.021 & -0.961 &  0.000 \\ 
     $z$ &  0.000 &  0.000 & -0.118 \\
    \vspace{-0.3cm} \\
    \hline \hline
    \multicolumn{4}{c}{$Z^*_{\mathrm{H}}$} \vspace{0.05cm} \\
    \hline 
     & $x$ & $y$ & $z$ \\
     $x$ &  0.031 & -0.124 & 0.000 \\ 
     $y$ & -0.003 &  0.015 & 0.000 \\ 
     $z$ &  0.000 &  0.000 & 0.143 \\
    \hline \hline
    \end{tabular}
    \end{minipage}
\end{table}

\subsection*{Born effective charges of PMI}
\vspace{-1cm}
\begin{table}[!htb]
    \centering
    \caption{For three representative values of the long-range mixing parameter $c_{\mathrm{LR}}$ (including the critical value where the chain dimerization occurs), the components of the effective charge tensors for each atom of the unit cell of PMI are reported. The $Z^*_{xx}$ component of the backbone atoms reaches strikingly high values, hint of the anomalous polar response of the $\pi$-orbitals along the backbone chain. The sum rule $Z^*_{\mathrm{C}} + Z^*_{\mathrm{N}} + Z^*_{\mathrm{H}} = 0$ is also respected.\vspace{.4cm}}
    \begin{minipage}[c]{0.3\textwidth}
    \centering
    \begin{tabular}{cccc}
    \multicolumn{4}{c}{$\mathbf{c_\mathrm{\textbf{LR}} \boldsymbol{=} 1\boldsymbol{\%}}$} \\
    \hline \hline
    \multicolumn{4}{c}{$Z^*_{\mathrm{C}}$} \vspace{0.05cm} \\
    \hline 
    & $x$ & $y$ & $z$ \\
    $x$ &  14.063 & 0.000 & 0.000 \\ 
    $y$ & -0.005  & 0.643 & 0.000 \\ 
    $z$ &  0.000  & 0.000 & 0.168 \\ 
    \vspace{-0.3cm} \\
    \hline \hline
    \multicolumn{4}{c}{$Z^*_{\mathrm{N}}$} \vspace{0.05cm} \\
    \hline 
    & $x$ & $y$ & $z$ \\
    $x$ & -14.059 &  0.000 &  0.000 \\ 
    $y$ &  0.005  & -0.427 &  0.000 \\ 
    $z$ &  0.000  &  0.000 & -0.273 \\
    \vspace{-0.3cm} \\
    \hline \hline
    \multicolumn{4}{c}{$Z^*_{\mathrm{H}}$} \vspace{0.05cm} \\
    \hline 
    & $x$ & $y$ & $z$ \\
    $x$ & -0.004 &  0.000 & 0.000 \\ 
    $y$ &  0.000 & -0.216 & 0.000 \\ 
    $z$ &  0.000 &  0.000 & 0.105 \\
    \hline \hline
    \end{tabular}
    \end{minipage}
    \begin{minipage}[c]{0.3\textwidth}
    \centering
    \begin{tabular}{cccc}
    \multicolumn{4}{c}{$\boldsymbol{c_\mathrm{\textbf{LR}} = 25\% \, (\simeq \overline{c}_\mathrm{\textbf{LR}})}$} \\
    \hline \hline
    \multicolumn{4}{c}{$Z^*_{\mathrm{C}}$} \vspace{0.05cm} \\
    \hline 
    & $x$ & $y$ & $z$ \\
    $x$ & 15.204 & 0.004 & 0.000 \\ 
    $y$ & -0.012 & 0.655 & 0.000 \\ 
    $z$ &  0.000 & 0.000 & 0.186 \\
    \vspace{-0.3cm} \\
    \hline \hline
    \multicolumn{4}{c}{$Z^*_{\mathrm{N}}$} \vspace{0.05cm} \\
    \hline
    & $x$ & $y$ & $z$ \\
    $x$ & -15.217 & -0.001 &  0.000 \\ 
    $y$ &  0.012  & -0.448 &  0.000 \\ 
    $z$ &  0.000  &  0.000 & -0.292 \\ 
    \vspace{-0.3cm} \\
    \hline \hline
    \multicolumn{4}{c}{$Z^*_{\mathrm{H}}$} \vspace{0.05cm} \\
    \hline
    & $x$ & $y$ & $z$ \\
    $x$ & 0.013 & -0.003 & 0.000 \\ 
    $y$ & 0.000 & -0.206 & 0.000 \\ 
    $z$ & 0.000 &  0.000 & 0.106 \\
    \hline \hline
    \end{tabular}
    \end{minipage}
    \begin{minipage}[c]{0.3\textwidth}
    \centering
    \begin{tabular}{cccc}
    \multicolumn{4}{c}{$\mathbf{c_\mathrm{\textbf{LR}} \boldsymbol{=} 100\boldsymbol{\%}}$} \\
    \hline \hline
    \multicolumn{4}{c}{$Z^*_{\mathrm{C}}$} \vspace{0.05cm} \\
    \hline 
    & $x$ & $y$ & $z$ \\
    $x$ &  5.762 & 0.166 & 0.000 \\ 
    $y$ & -0.268 & 0.638 & 0.000 \\ 
    $z$ &  0.000 & 0.000 & 0.203 \\
    \vspace{-0.3cm} \\
    \hline \hline
    \multicolumn{4}{c}{$Z^*_{\mathrm{N}}$} \vspace{0.05cm} \\
    \hline 
    & $x$ & $y$ & $z$ \\
    $x$ & -5.699 & 0.022 &  0.000 \\ 
    $y$ & 0.267 & -0.476 &  0.000 \\ 
    $z$ &  0.000 &  0.000 & -0.318 \\   
    \vspace{-0.3cm} \\
    \hline \hline
    \multicolumn{4}{c}{$Z^*_{\mathrm{H}}$} \vspace{0.05cm} \\
    \hline 
    & $x$ & $y$ & $z$ \\
    $x$ & -0.063 & -0.188 & 0.000 \\ 
    $y$ &  0.001 & -0.162 & 0.000 \\ 
    $z$ &  0.000 &  0.000 & 0.115 \\
    \hline \hline
    \end{tabular}
    \end{minipage}
    \label{tab:Zeff_PMI}
\end{table}

\clearpage

\subsection*{Born effective charges of PVDF}
\vspace{-1cm}
\begin{table}[!htb]
    \centering
    \caption{Components of the effective charge tensors of the atoms of the unit cell of PVDF. For each component, values reported are obtained as average of those computed with different values of $c_{\mathrm{LR}}$, the highest standard deviation being of the order of $10^{-2}$. The very weak dependence of the effective charges of PVDF on $c_{\mathrm{LR}}$ is a direct consequence of the fact that the structure of PVDF does not depend on $c_{\mathrm{LR}}$, as observed in Figure \ref{fig:PVDF_structure}. Our results for the Born effective charges of PVDF are overall consistent with previously reported values calculated within the generalized-gradient approximation\cite{RAMER200510431}. The sum rule $Z^*_{\mathrm{C_F}} + Z^*_{\mathrm{C_H}} + Z^*_{\mathrm{F_1}} + Z^*_{\mathrm{F_2}} + Z^*_{\mathrm{H_1}} + Z^*_{\mathrm{H_2}} = 0$ is also respected.\vspace{.4cm}} 
    \begin{minipage}[c]{0.3\textwidth}
    \centering
    \begin{tabular}{cccc}
    \hline \hline
    \multicolumn{4}{c}{$Z^*_{\mathrm{C_H}}$} \vspace{0.05cm} \\
    \hline 
    & $x$ & $y$ & $z$ \\
    $x$ & -0.603 &  0.000 & 0.000 \\ 
    $y$ &  0.000 & 0.012 &  0.000 \\ 
    $z$ & -0.001 & 0.001 & -0.092 \\
    \vspace{-0.3cm} \\
    \hline \hline
    \multicolumn{4}{c}{$Z^*_{\mathrm{C_F}}$} \vspace{0.05cm} \\
    \hline 
    & $x$ & $y$ & $z$ \\
    $x$ & 1.333 & 0.000 &  0.000 \\ 
    $y$ & 0.000 & 1.165 & 0.000 \\ 
    $z$ & 0.001 &  0.000 & 0.955 \\
    \vspace{-0.3cm} \\
    \hline \hline
    \end{tabular}
    \end{minipage}
    \hspace{-0.75cm}
    \begin{minipage}[c]{0.3\textwidth}
    \centering
    \begin{tabular}{cccc}
    \hline \hline
    \multicolumn{4}{c}{$Z^*_{\mathrm{H_1}}$} \vspace{0.05cm} \\
    \hline 
    & $x$ & $y$ & $z$ \\
    $x$ & 0.069 & 0.000 & 0.000 \\ 
    $y$ &  0.000 & 0.002 & -0.048 \\ 
    $z$ &  0.000 & -0.060 & 0.034 \\
    \vspace{-0.3cm} \\
    \hline \hline
    \multicolumn{4}{c}{$Z^*_{\mathrm{F_1}}$} \vspace{0.05cm} \\
    \hline 
    & $x$ & $y$ & $z$ \\
    $x$ & -0.434 &  0.000 &  0.000 \\ 
    $y$ & -0.002 & -0.590 & -0.292 \\ 
    $z$ & 0.000 & -0.248 & -0.467 \\
    \vspace{-0.3cm} \\
    \hline \hline
    \end{tabular}
    \end{minipage}
    \hspace{-0.75cm}
    \begin{minipage}[c]{0.3\textwidth}
    \centering
    \begin{tabular}{cccc}
    \hline \hline
    \multicolumn{4}{c}{$Z^*_{\mathrm{H_2}}$} \vspace{0.05cm} \\
    \hline 
    & $x$ & $y$ & $z$ \\
    $x$ & 0.069 & 0.000 &  0.000 \\ 
    $y$ & 0.000 & 0.003 & 0.048 \\ 
    $z$ &  0.000 & 0.060 & 0.035 \\
    \vspace{-0.3cm} \\
    \hline \hline
    \multicolumn{4}{c}{$Z^*_{\mathrm{F_2}}$} \vspace{0.05cm} \\
    \hline 
    & $x$ & $y$ & $z$ \\
    $x$ & -0.434 &  0.000 & 0.000 \\ 
    $y$ & 0.002 & -0.592 & 0.292 \\ 
    $z$ & 0.000 & 0.246 & -0.465 \\
    \vspace{-0.3cm} \\
    \hline \hline    
    \end{tabular}
    \end{minipage}
    \label{tab:Zeff_PVDF}
\end{table}

\subsection*{Piezoelectricity of packed polymers}

\begin{figure}[!htb]
    \begin{minipage}[c]{1.0\linewidth}
    \centering
    \includegraphics[width=0.45\textwidth]{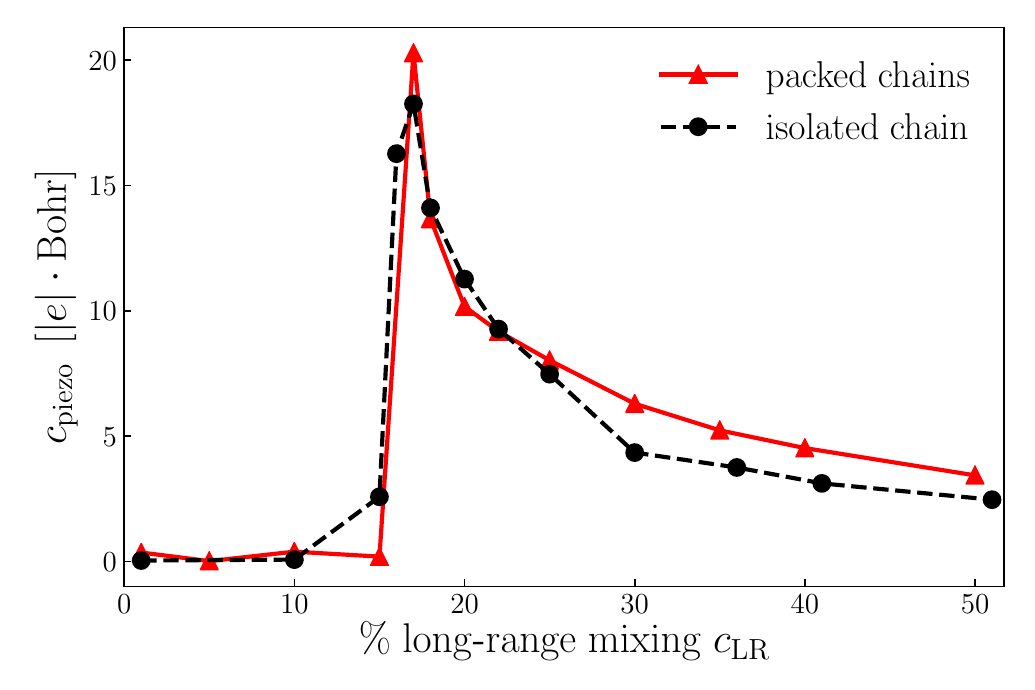}
    \includegraphics[width=0.45\textwidth]{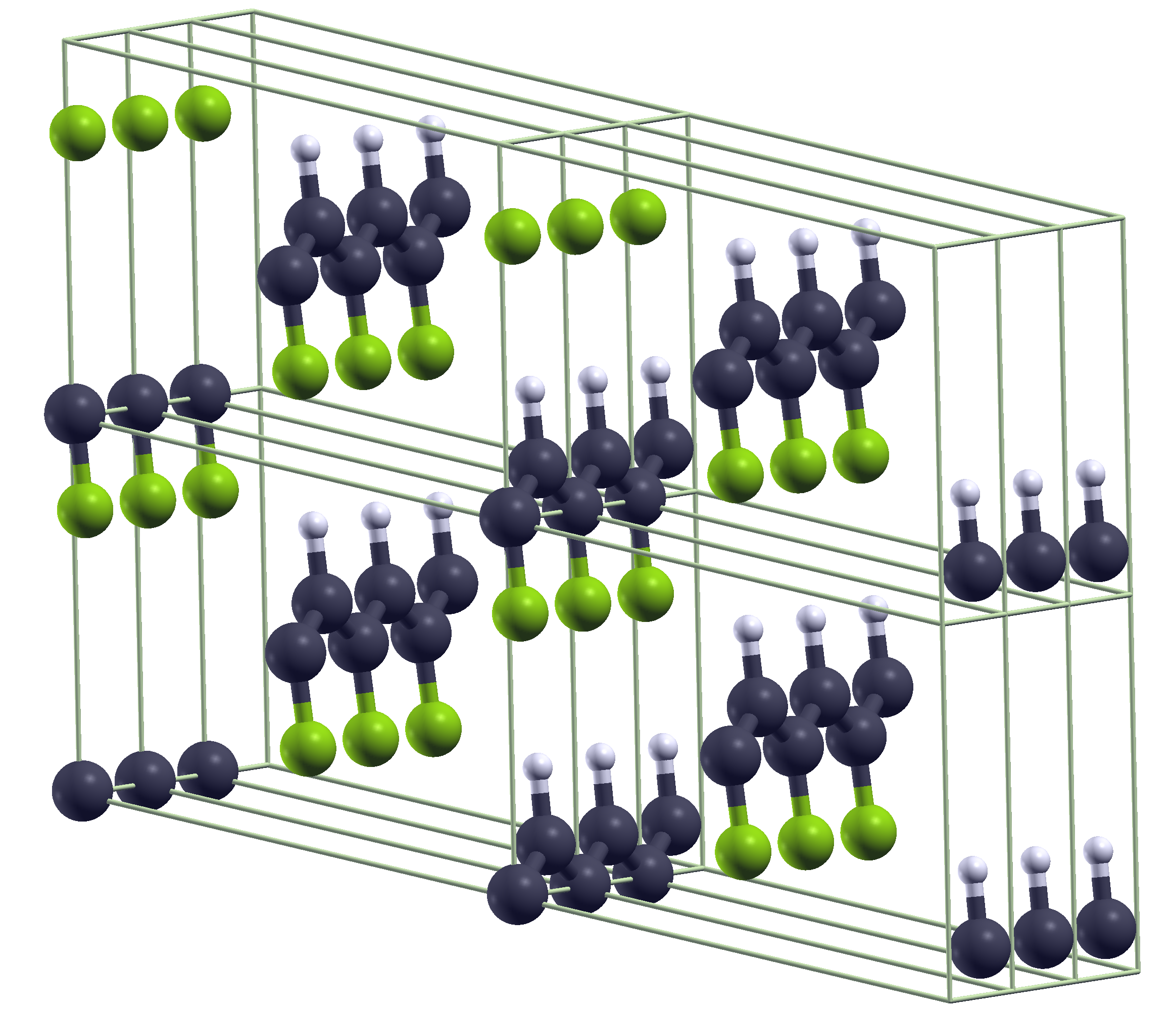}
    \end{minipage}
    \caption{On the left panel, piezoelectric coefficients of MFPA in its isolated polymeric chain configuration are compared to those obtained for packed chains. The 3D structure adopted, displayed on the right panel, is similar to the one adopted for PVDF in Ref.\cite{nakhmanson2004ab}. The agreement between the results highlight the robustness of the enhancement mechanism.}
    \label{fig:MFPA_bulk}
\end{figure}

\end{document}